\documentclass[aps,prl,reprint,superscriptaddress]{revtex4-1} 

\usepackage{graphicx}
\usepackage{amsmath}
\usepackage{amssymb}
\usepackage{siunitx}
\usepackage{url}

\newcommand{\ensavg}[1]{\left< #1 \right>} 
\newcommand{\ket}[1]{\left\lvert #1 \right\rangle}%
 
\begin{document}

\title{Bad metallic transport in a cold atom Fermi-Hubbard system}

\author{Peter T. Brown}
\affiliation{Department of Physics, Princeton University, Princeton, New Jersey 08544 USA}
\author{Debayan Mitra}
\affiliation{Department of Physics, Princeton University, Princeton, New Jersey 08544 USA}
\author{Elmer Guardado-Sanchez}
\affiliation{Department of Physics, Princeton University, Princeton, New Jersey 08544 USA}
\author{Reza Nourafkan}
\affiliation{D\'epartement de Physique, Institut Quantique, and Regroupement Qu\'eb\'ecois sur les Mat\'eriaux de Pointe, Universit\'e de Sherbrooke, Sherbrooke, Qu\'ebec, Canada J1K 2R1}
\author{Alexis Reymbaut}
\affiliation{D\'epartement de Physique, Institut Quantique, and Regroupement Qu\'eb\'ecois sur les Mat\'eriaux de Pointe, Universit\'e de Sherbrooke, Sherbrooke, Qu\'ebec, Canada J1K 2R1}
\author{Charles-David H\'ebert}
\affiliation{D\'epartement de Physique, Institut Quantique, and Regroupement Qu\'eb\'ecois sur les Mat\'eriaux de Pointe, Universit\'e de Sherbrooke, Sherbrooke, Qu\'ebec, Canada J1K 2R1}
\author{Simon Bergeron}
\affiliation{D\'epartement de Physique, Institut Quantique, and Regroupement Qu\'eb\'ecois sur les Mat\'eriaux de Pointe, Universit\'e de Sherbrooke, Sherbrooke, Qu\'ebec, Canada J1K 2R1}
\author{A.–-M.  S. Tremblay}
\affiliation{D\'epartement de Physique, Institut Quantique, and Regroupement Qu\'eb\'ecois sur les Mat\'eriaux de Pointe, Universit\'e de Sherbrooke, Sherbrooke, Qu\'ebec, Canada J1K 2R1}
\affiliation{Canadian Institute for Advanced Research, Toronto, Ontario, Canada M5G 1Z8}
\author{Jure Kokalj}
\affiliation{Faculty of Civil and Geodetic Engineering, University of Ljubljana, SI-1000 Ljubljana, Slovenia}
\affiliation{Jo\v zef Stefan Institute, Jamova 39, SI-1000 Ljubljana, Slovenia}
\author{David A. Huse}
\affiliation{Department of Physics, Princeton University, Princeton, New Jersey 08544 USA}
\author{Peter Schau{\ss}}
\affiliation{Department of Physics, Princeton University, Princeton, New Jersey 08544 USA}
\affiliation{Present address: Department of Physics, University of Virginia, Charlottesville, Virginia 22904 USA}
\author{Waseem S. Bakr}
\email{wbakr@princeton.edu}
\affiliation{Department of Physics, Princeton University, Princeton, New Jersey 08544 USA}

\date{\today}

\begin{abstract}
Strong interactions in many-body quantum systems complicate the interpretation of charge transport in such materials. To shed light on this problem, we study transport in a clean quantum system: ultracold $^6$Li in a 2D optical lattice, a testing ground for strong interaction physics in the Fermi-Hubbard model. We determine the diffusion constant by measuring the relaxation of an imposed density modulation and modeling its decay hydrodynamically. The diffusion constant is converted to a resistivity using the Nernst-Einstein relation. That resistivity exhibits a linear temperature dependence and shows no evidence of saturation, two characteristic signatures of a bad metal. The techniques we develop here may be applied to measurements of other transport quantities, including the optical conductivity and thermopower.
\end{abstract}

\maketitle


In conventional materials, charge is carried by quasiparticles and conductivity is understood as a current of these charge carriers developed in response to an external field. For the conductivity to be finite, the charge carriers must be able to relax their momentum through scattering. The Boltzmann kinetic equation in conjunction with Fermi liquid theory provides a detailed description of transport in conventional materials, including two trademarks of resistivity. The first is the Fermi liquid prediction that the temperature-dependent resistivity $\rho(T)$ should scale like $T^2$ at low temperature \cite{Coleman2015}. The second is that the resistivity should not exceed a maximum value $\rho_{\text{max}}$, obtained from the Drude relation assuming the Mott-Ioffe-Regel (MIR) limit which states that the mean free path of a quasiparticle cannot be less than the lattice spacing \cite{Ioffe1960, Mott1972}. This resistivity bound itself is sometimes referred to as the MIR limit.

Strong interactions can however lead to a breakdown of Fermi liquid theory. One signal of this breakdown is anomalous scaling of $\rho$ with temperature, including the linear scaling observed in the strange metal'' state of the cuprates \cite{Hussey2008} and other anomalous scalings in d- and f- electron materials \cite{Stewart2001}. Another is the violation of the resistivity bound $\rho < \rho_{\text{max}}$, which is observed in a wide variety of materials \cite{Gunnarsson2003}. Additionally, interactions may lead to a situation where the momentum relaxation rate alone does not determine the conductivity, in contrast to the semiclassical Drude formula, generalizations of which hold for a large class of systems called coherent metals \cite{Hartnoll2014}. Approaches introduced to understand these anomalous behaviors include hidden Fermi liquids \cite{Anderson2008}, marginal Fermi liquids \cite{Varma1989}, proximity to quantum critical points \cite{Vucicevic2015} and associated holographic approaches \cite{Hartnoll2018}, and many numerical studies of model systems, most notably the Hubbard \cite{Scalapino2007} and $t-J$ \cite{Jaklic2000} models.

Disentangling strong interaction physics from other effects, such as impurities and electron-phonon coupling, is difficult in real materials. Cold atom systems are free of these complications, but transport experiments are challenging due to the finite and isolated nature of these systems. Most fermionic charge transport experiments have focused on either studying mass flow through optically structured mesoscopic devices \cite{Brantut2012, Valtolina2015, Krinner2017, Lebrat2017} or bulk transport in lattice systems \cite{Ott2004, Strohmaier2007, Schneider2012, Xu2016, Anderson2017}. Here, we explore bulk transport in a Fermi-Hubbard system by studying charge diffusion, which is a microscopic process related to conductivity through the Nernst-Einstein equation $\sigma = \chi_c D$, where $D$ is the diffusion constant and $\chi_c = \left. \left(\frac{\partial n}{\partial \mu}\right) \right\vert_T$ is the compressibility. This relation requires only the assumption of linear response and absence of thermoelectric coupling \cite{supplement} and does not rest on assumptions concerning quasiparticles.


\begin{figure*}[ht]
\includegraphics[width=\textwidth]{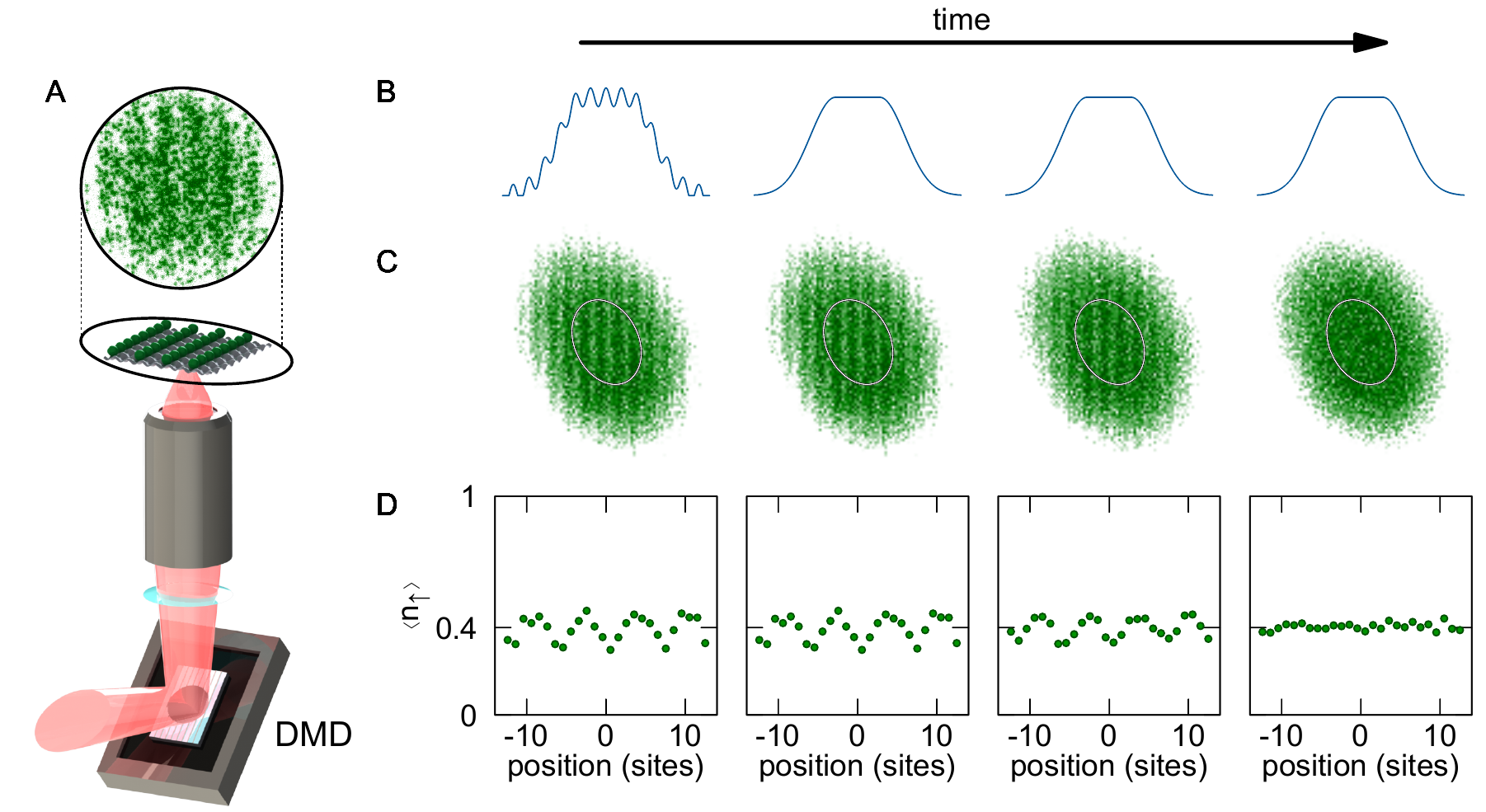}
\caption{{\bfseries Measuring transport in the Hubbard model.} \label{fig:schematic} ({\bfseries A}) Top: Exemplary single shot fluorescence image of the atomic density for one spin component. Field of view diameter is approximately $60 \text{ $\mu$m}$. Bottom: Schematic of the setup for generating optical potentials. Far-off-resonant light is projected onto a digital micromirror device (DMD) and the resulting pattern is imaged onto the atoms using a high-resolution objective. We project a sinusoidally modulated potential along one direction. ({\bfseries B}) One dimensional cuts along the projected potential. The DMD is used to flatten the trap and project a sinusoidally modulated potential (leftmost image). The confining potential comes from the optical lattice. After initial preparation, the sinusoidal potential is suddenly turned off, but the flattening potential is not. ({\bfseries C}) Average density of a single spin component, $\ensavg{n_\uparrow}$, versus time for approximately $30$ images. Initially the system is in thermal equilibrium with a spatially modulated density (leftmost image, $ 0 \text{$\mu$s}$ decay time). Immediately after the sinusoidal potential is turned off, the system is no longer in equilibrium but the density has not yet changed (second from left, $ 0 \text{$\mu$s}$ decay time). The density modulation decays with time (third from left, $ 50 \text{$\mu$s}$ decay time) until it is no longer visible (fourth from left, $500 \text{$\mu$s}$ decay time). The central flattened region of the potential is marked by a white ellipse. The field of view is approximately $75 \text{ $\mu$m} \times 75 \text{ $\mu$m}$. ({\bfseries D}) Atomic density from C averaged along the direction orthogonal to the modulation in the central flattened region of the potential.} 
\end{figure*}


We realize the 2D Fermi Hubbard model using a degenerate spin-balanced mixture of two hyperfine ground states of $^6$Li in an optical lattice \cite{Brown2017}. Our lattice beams produce a harmonic trapping potential, which leads to a varying atomic density in the trap. To obtain a system with uniform density, we flatten our trapping potential over an elliptical region of mean diameter 30 sites using a repulsive potential created with a spatial light modulator. We superimpose an additional sinusoidal potential that varies slowly along one direction of the lattice with a controllable wavelength (Fig.~\ref{fig:schematic}, A and B). By adiabatically loading the gas into these potentials, we prepare a Hubbard system in thermal equilibrium with a small amplitude (typically 10\%)  sinusoidal density modulation. The average density in the region with the flattened potential is the same with and without the sinusoidal potential. Next, we suddenly turn off the added sinusoidal potential and observe the decay of the density pattern versus time (Fig.~\ref{fig:schematic}, C and D), always keeping the optical lattice at fixed intensity. We measure the density of a single spin component, $\ensavg{n_\uparrow}$ using techniques described in \cite{Brown2017}, giving us access to the total density through $\ensavg{n}=2\ensavg{n_\uparrow}$. 

We work at average total density $\ensavg{n} = 0.82(2)$. This value is close to a conjectured quantum critical point in the Hubbard model \cite{Vidhyadhiraja2009}. Our lattice depth is 6.9(2) $E_R$, where $E_R/h = 14.66$ kHz is the lattice recoil, leading to a tunneling rate of $t/h = 925(10)\text{ Hz}$. Here $h$ is Planck's constant. We adjust the scattering length, $a_s = 1070(10) a_o$, by working at a magnetic bias field of \SI{616.0(2)}{G}, in the vicinity of the Feshbach resonance centered near \SI{690}{G}. These parameters lead to an on-site interaction to tunneling ratio $U/t = 7.4(8)$, which is in the strong-interaction regime and close to the value that maximizes antiferromagnetic correlations at half-filling \cite{Khatami2011}.


\begin{figure*}[ht]
\includegraphics[width=\textwidth]{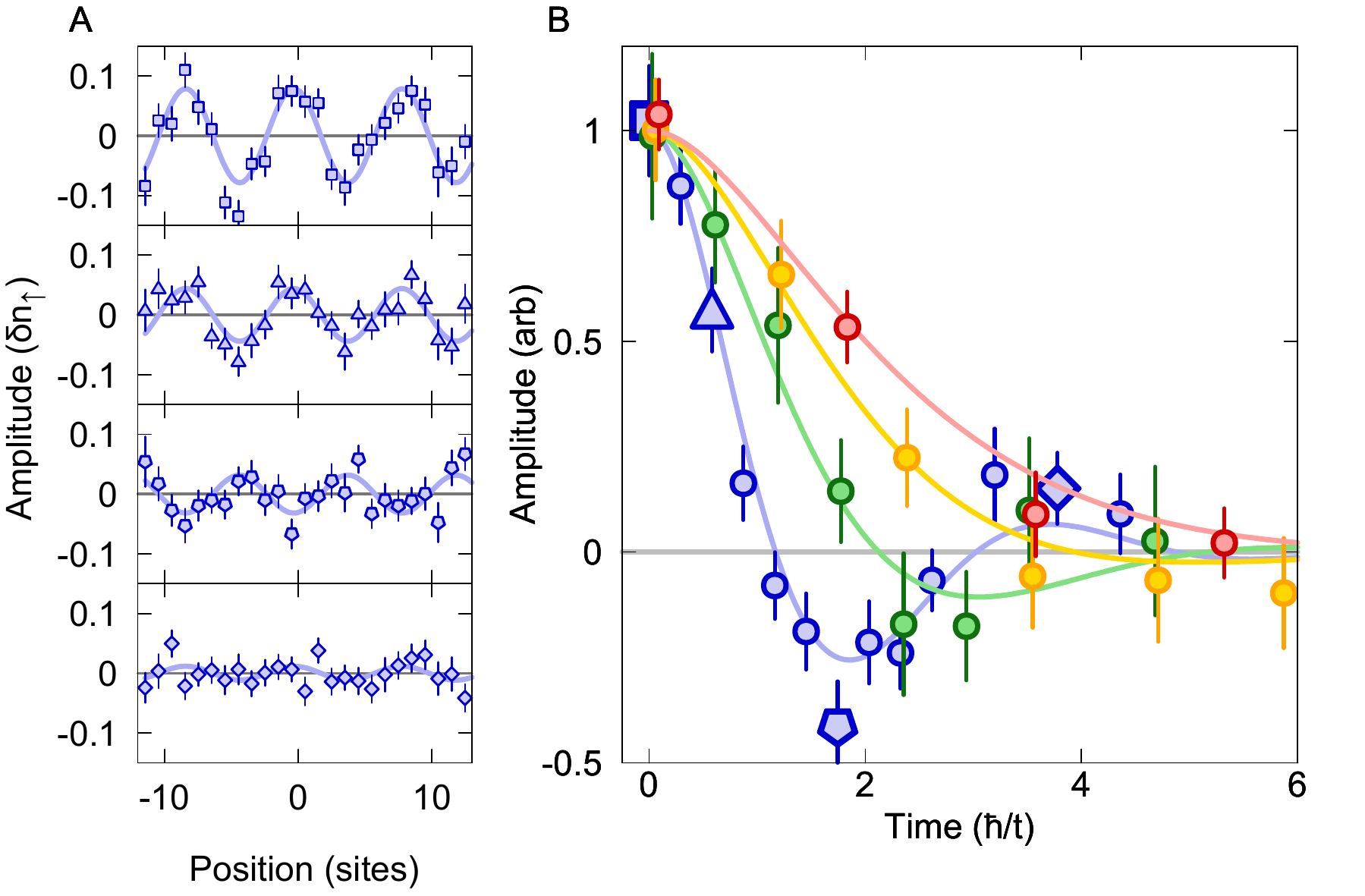}
\caption{{\bfseries Decay of modulation pattern versus time.} \label{fig:densitymod} ({\bfseries A}) Cloud profiles averaged along the direction of the modulation (points) and sinusoidal fits (lines) for modulation wavelength 8.1 sites and times $0 \ \hbar/t$ (top), $0.6 \ \hbar/t$ (second from top), $1.7 \ \hbar/t$ (third from top), and $3.8 \ \hbar/t$ (bottom). The average value obtained from the sine fit has been subtracted. ({\bfseries B}) Sinusoid fit amplitudes (points) versus decay time for modulation periods 8.1 (blue), 11.8 (green), 15.6 (yellow), and 18.7 (red) sites. Each curve is scaled by the initial modulation amplitude. Lines are obtained from a simultaneous fit of the diffusion constant, $D$, and momentum relaxation rate, $\Gamma$, to all wavelengths and times \cite{supplement}. Different shaped points for period 8.1 correspond to different panels in A, $0 \ \hbar/t$ (square), $0.6 \ \hbar/t$ (triangle), $1.7 \ \hbar/t$ (pentagon), and $3.8 \ \hbar/t$ (diamond). The temperature for all wavelengths and decay times is $T/t = 0.57(8)$. Each point is the average of approximately $30$ images. Error bars standard error of the mean.}
\end{figure*}


We observe the decay of the initial sinusoidal density pattern over a period of a few tunneling times. The short timescale ensures that the observed dynamics are not affected by the inhomogeneous density outside of the central flattened region of the trap. To obtain better statistics, we apply the sinusoidal modulation along one dimension and average along the other direction (Fig.~\ref{fig:schematic}, A and C). We fit the average modulation profile to a sinusoid, where the phase and frequency are fixed by the initial pattern (Fig.~\ref{fig:densitymod}A). The time dependence of the amplitude of the sinusoid quantifies the decay of the density modulation, Fig.~\ref{fig:densitymod}B. Our experimental technique is analogous to that of \cite{Hild2014}, which studied the decay of a sinusoidally modulated spin pattern in a bosonic system.

The decay of the sinusoidal density pattern versus the wavelength of the modulation becomes consistent with diffusive transport at long wavelengths. In diffusive transport, the amplitude of a density pattern at wave vector $k = 2\pi/\lambda$ will decay exponentially with time constant $\tau = 1/Dk^2$, where $D$ is the diffusion constant. We observe exponentially decaying amplitudes with diffusive scaling for wavelengths longer than 15 sites. However, the decay curves are flat at early times, showing clear deviation from exponential decay. For short wavelengths, we observe deviations from diffusive behavior in the form of underdamped oscillations, which can be understood as the damped limit of sound waves. Both of these effects are related to the fact that a density modulation does not instantaneously create a current, as implied by the diffusion equation. Rather, a current requires a finite amount of time to reach an equilibrium value after the creation of a density modulation. 

To unify the description of modulation decay at all wavelengths, we developed a hydrodynamic description that conserves density and has a finite momentum (or current) relaxation rate \cite{supplement}. This approach leads to a differential equation for the density decay,
\begin{equation*}
\partial_t^2 n + \Gamma \partial_t n + \Gamma D k^2 n = 0,
\end{equation*}
where $\Gamma$ is the momentum-relaxation rate and $D$ is the diffusion constant. This oscillator model crosses over from an underdamped to an overdamped (approximately diffusive) regime at a modulation wavelength $4\pi \sqrt{D/\Gamma}$. Instead of assuming that $D$ and $\Gamma$ are dependent parameters linked through a Drude formula, as would be the case in a system that can be described using quasiparticles, we determine $D$ and $\Gamma$ from our data at a fixed temperature, simultaneously fitting the amplitude as a function of time for all wavelengths as shown in Fig.~\ref{fig:densitymod}B. 

Our model neglects thermoelectric effects, which affect the measured density response by coupling local energy density modulations, and the resulting temperature gradients, to the particle current. We justify this approximation based on the empirical fact that our simple model fits the data and that we have not been able to detect any measurable temperature modulation in the gas \cite{supplement}. In addition, theoretical work suggests that the thermopower (Seebeck coefficient) is negligible near our doping \cite{Palsson1998, Jaklic2000}. 


\begin{figure}
\includegraphics[width=\columnwidth]{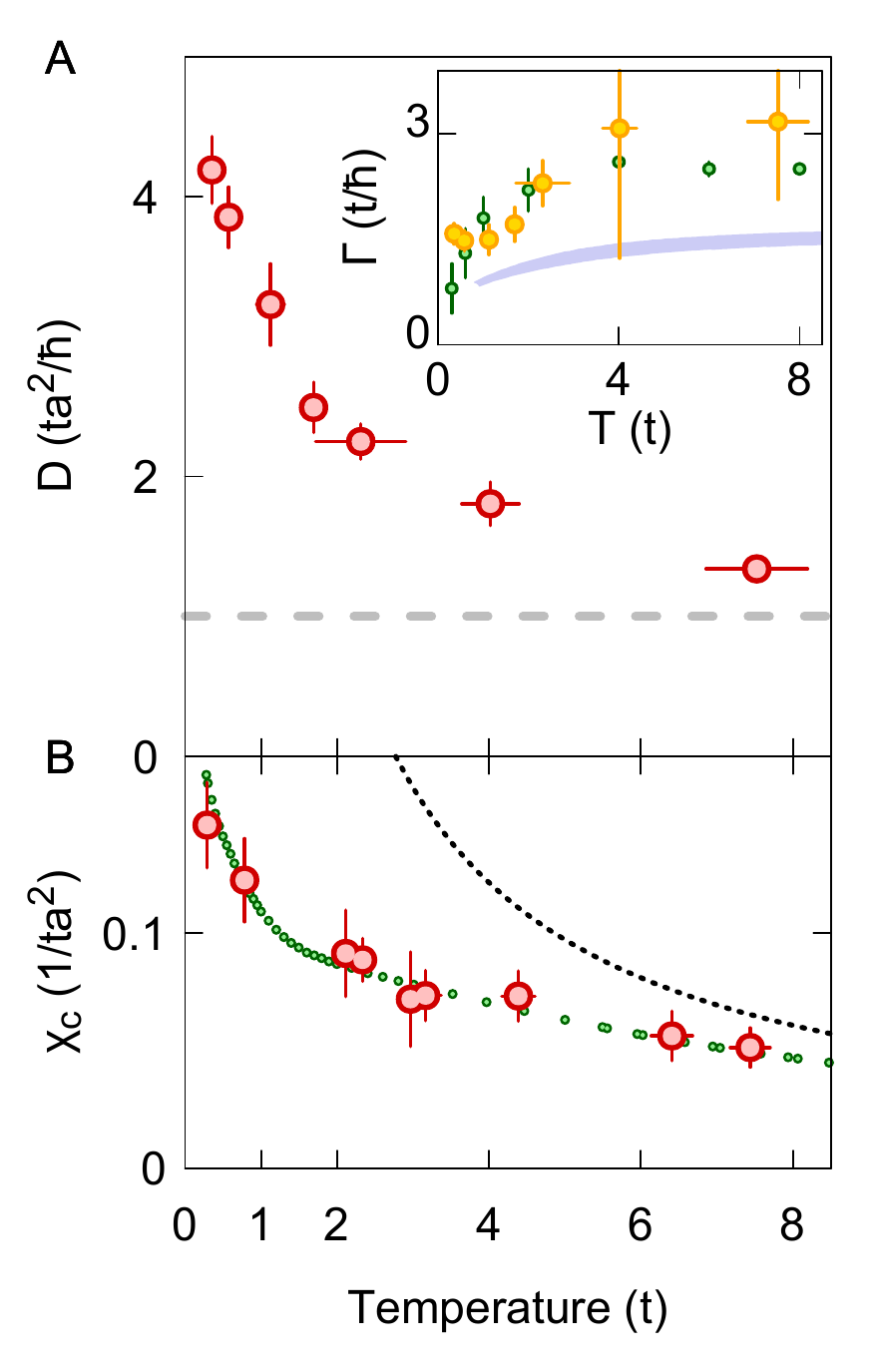}
\caption{{\bfseries Hydrodynamic model parameters.} \label{fig:hydromodelparams} ({\bfseries A}) Experimental diffusion constant, $D$, versus temperature (red) and the lower bound on $D$ inferred from the Mott-Ioffe-Regel limit (grey). Each point is typically determined from $4$ different modulation wavelengths each consisting of $10$ different decay times with $30$ images for each decay time.  ({\bfseries Inset}) Results for the momentum relaxation rate, $\Gamma$, including experimental data (yellow), single-site dynamical mean-field theory results for $\ensavg{n} = 0.825$ and $U/t = 7.5$ (green), and finite-temperature Lanczos method results on a 16-site cluster for $\ensavg{n} = 0.8-0.85$ and $U/t = 7.5$ (blue band). ({\bfseries B}) Results for the charge compressibility, $\chi_c$. Experimental results (red points), determinantal quantum Monte Carlo at $\ensavg{n} = 0.83$ and $U/t = 7.5$ (green points), and the high-temperature limit $1/T$ scaling (black dashed line). Each compressibility point is typically determined from $60$ images. Experimental error bars standard error of the mean.}
\end{figure}


The temperature dependence of $D$ and $\Gamma$ are the focus of the rest of this paper. The temperature is controlled as follows. After the initial preparation of the cloud, we hold the atoms in the trap or modulate the lattice amplitude for a controlled time to heat the system. To determine the temperature of the cloud after the system has equilibrated, we measure the singles density or local moment, $\ensavg{n^s} = \ensavg{n_\uparrow + n_\downarrow - 2n_\uparrow n_\downarrow}$, and the nearest-neighbor correlations between spin-up atoms $C^\uparrow(\mathbf{d}) = 4 \left(\ensavg{n_{\mathbf{i} + \mathbf{d},\uparrow} n_{\mathbf{i}\uparrow}} - \ensavg{n_{\mathbf{i} + \mathbf{d},\uparrow}} \ensavg{n_{\mathbf{i}\uparrow}} \right)$, where $\mathbf{i} = (i_x, i_y)$. We compare these quantities to determinantal quantum Monte Carlo (DQMC) simulations to extract the temperature \cite{supplement}. For temperatures at the low end of the range we can access, between $0.3 < T/t < 1 $, the density correlations are a sensitive thermometer. At higher temperature the singles density becomes a better thermometer. We have compared the temperature of the gas before switching off the potential modulation and after the density modulation has decayed, and find no measurable increase.

As the temperature is lowered, Pauli blocking closes scattering channels, leading to an increased range of diffusion, in agreement with our observations in Fig.~\ref{fig:hydromodelparams}A. At high temperatures, $D$ is expected to saturate, eventually approaching an infinite temperature limiting value \cite{Perepelitsky2016}. The diffusion constant is closely related to the mean-free-path, $l$, and is often estimated as $D = l \ensavg{v} /2$, where $\ensavg{v}$ is the mean quasiparticle velocity \cite{Kokalj2017}. Therefore, the MIR limit implies a lower bound on the diffusion constant, $D \gtrsim t a^2 / \hbar$, where $a$ is the lattice constant. Our measured diffusion constants approach this derived bound at high temperatures, but do not violate it. Because of the difficulty of measuring diffusion constants in materials, this limit has not been tested in real bad metals. We do not compare the measured diffusion constants with theory because determining $D$ requires working in the limit $\lambda \rightarrow \infty$ \cite{supplement}, and exact techniques such as diagonalization of finite systems and DQMC are limited to small system sizes. Even determining the infinite temperature limiting value is a non-trivial quantum dynamics problem \cite{Mukerjee2006, Leviatan2017}. 

In a clean system like ours momentum relaxation can only occur thanks to umklapp scattering, where a portion of the net momentum in a collision is transferred to the rigid lattice. Nevertheless, the momentum relaxation is strong at our interaction strength which makes determining the temperature dependence of $\Gamma$ challenging because $\Gamma$ drops out of the model entirely in the overdamped limit. We find that $\Gamma$ decreases weakly with decreasing temperature (Fig.~\ref{fig:hydromodelparams}, Inset). This trend may again be understood as Pauli-blocking suppressing momentum relaxation at low temperatures.

We compare the experimental $\Gamma$ to results from state-of-the-art finite-temperature Lanczos method (FTLM) and dynamical mean-field theory (DMFT) simulations by estimating the momentum relaxation rate as the half-width at half-maximum of the Drude peak in the optical conductivity. The optical conductivity has an additional peak at $\omega \sim U$, but this does not affect $\Gamma$ significantly \cite{supplement}. Our experimental $\Gamma$ agrees reasonably with the DMFT results, but exceeds the FTLM results by up to a factor of two. FTLM is an exact technique expected to give correct results at high temperature. One possible explanation for the discrepancy is that $\Gamma$ is sensitive to the amplitude of the density modulation. To test this, we measured $\Gamma$ and $D$ versus modulation amplitude \cite{supplement}. We found $D$ is insensitive to the amplitude in the range explored. $\Gamma$ shows some amplitude dependence but, because of the large error bars we can not conclusively say if this is the source of the discrepancy between experiment and FTLM (Fig.~S1).

 To extract a resistivity using the Nernst-Einstein relation, we need the compressibility. It is determined in a separate experiment by measuring the variation of total density versus position in a harmonic trap and converting the position to chemical potential in the local density approximation \cite{Ho2009, Cocchi2016, supplement}. The measured compressibility increases with decreasing temperature (Fig.~\ref{fig:hydromodelparams}B). For our highest experimental temperatures, $\chi_c$ approaches $n(1 - n/2)/T$, as expected in the high temperature limit \cite{Perepelitsky2016}. At sufficiently low temperature $\chi_c$ is expected to saturate, but we do not reach this limit at our lowest experimental temperature, $T/t = 0.3$. Our experimental results agree well with DQMC numerics over the full range of experimental temperatures.
 

\begin{figure*}
\includegraphics[width=\textwidth]{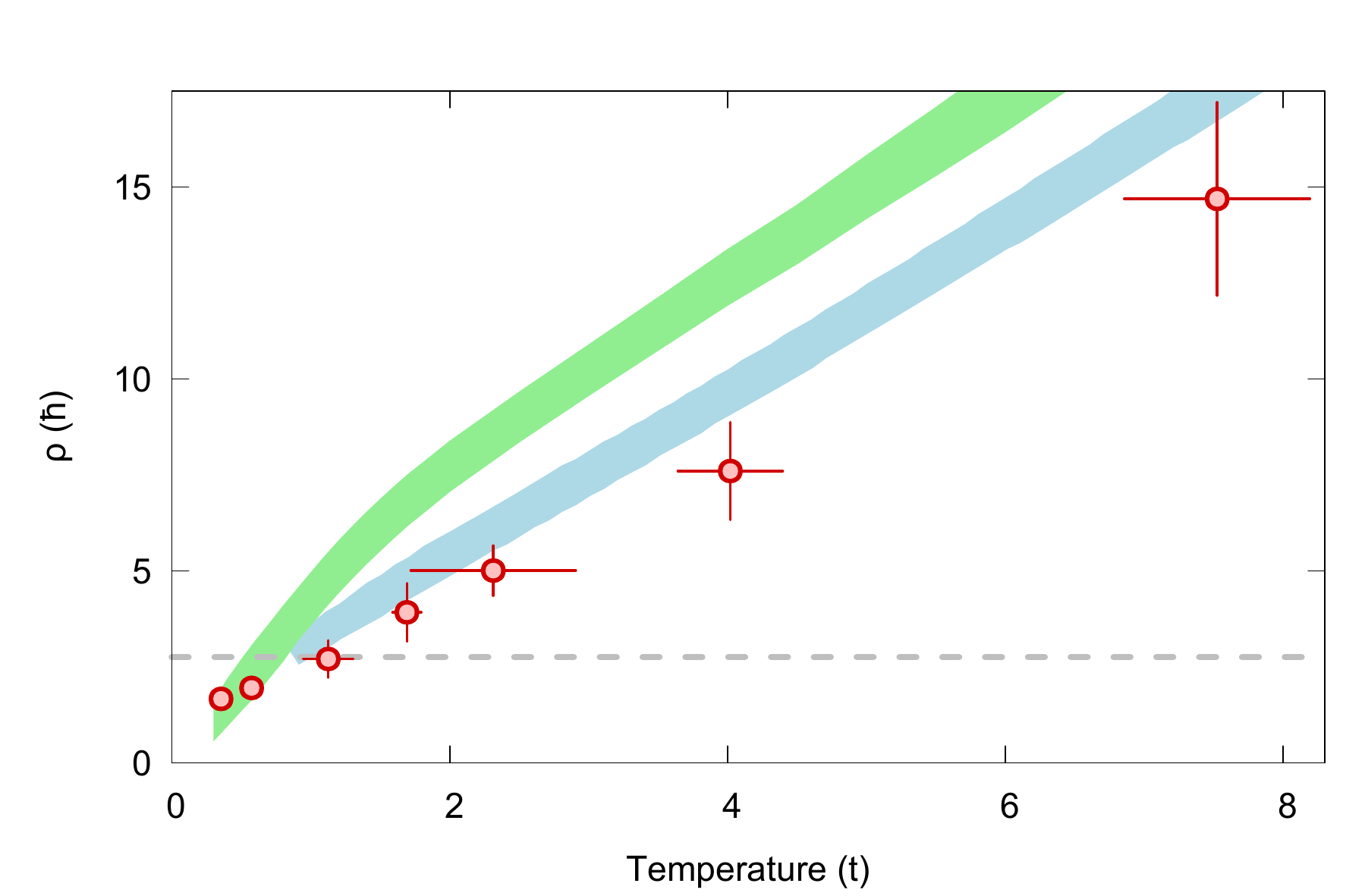}
\caption{{\bfseries Conductivity versus temperature.} \label{fig:conductivity} Results for the resistivity, $\rho$. Experiment (red), 16-site finite-temperature Lanczos method for $U/t = 7.5$ and $\ensavg{n}=0.8 - 0.85$ (light-blue band), single-site dynamical mean-field theory results for $U/t = 7.5$, $\ensavg{n}=0.825$ (green band), and the lower bound on conductivity inferred from the Drude relation using the Mott-Ioffe-Regel limit, $\rho_\text{max}$ (grey). For more information about the error bands, see \cite{supplement}.}
\end{figure*}
 

We can now use the Nernst-Einstein relation to determine the conductivity from the measured diffusion constant and charge compressibility. We examine the temperature dependence of the resistivity $\rho = 1/\sigma$ in Fig.~\ref{fig:conductivity}, and observe that it rises without limit, showing no sign of saturation. Assuming the existence of quasiparticles, the maximum resistivity obtained from the Drude relation using the MIR limit is $\rho < \rho_\text{max} \approx \sqrt{\frac{2 \pi}{n}} \hbar$ \cite{Gunnarsson2003, Kokalj2017}. We find that our resistivity violates this bound for temperatures above $T/t \sim 1.3$.  The temperature where $\rho$ exceeds this limit is near the Brinkman-Rice temperature scale, defined by $T_\text{BR} = (1-n) W$, where $W = 8t$ is the bandwidth, which is an estimate of the degeneracy temperature of quasiparticles in a doped Mott insulator. Similar violation of the resistivity bound at $T_\text{BR}$ has been observed in DMFT studies \cite{Deng2013, Xu2013}.

The failure of $\rho$ to saturate at the resistivity bound is similar to behavior observed in bad metals at high temperatures \cite{Gunnarsson2003}. In our system, the violation of the resistivity bound is not associated with the mean free path becoming shorter than the lattice spacing because the diffusion constant does not violate its derived bound, but rather with the temperature dependence of the compressibility \cite{Kokalj2017}. This suggests a need for a more careful distinction between the MIR limit on the mean free path and the resistivity bound, despite the presumed equivalence of these concepts in condensed matter experiments.

To further elucidate the temperature dependence of $\rho$, we fit our results to the form $\rho(T) = \rho_o + A T + B T^2$. We find the temperature dependence is linear to good approximation as we obtain $\rho_o = 1.1(1) \hbar$, $A = 1.55(15) \frac{\hbar}{t}$, and $B = 0.03(3) \frac{\hbar}{t^2}$. Alternatively, a power law fit to the form $\rho(T) = \rho_o + (C T)^\alpha$ yields $\rho_o = 1.2(2) \hbar$, $C = 1.4(2) \frac{\hbar}{t}$, and $\alpha = 1.1(1)$. Similar fits show the inverse diffusion constant $1/D$ scales with $\alpha = 0.6(1)$ and the inverse charge compressibility scales with $\alpha = 0.85(20)$. In our temperature range, the linear resistivity is a combined result of the temperature dependence of the diffusivity and compressibility, both of which behave in a non-trivial way. This behavior should be contrasted with the high-temperature regime, $T \gg W$, where $D$ saturates to a limiting value and the resistivity inherits its temperature dependence from the compressibility, which scales as $\chi_c \propto 1/T$  \cite{Perepelitsky2016}. It should also be contrasted with the low-temperature regime usually considered in condensed matter where the compressibility has saturated and the resistivity inherits its temperature dependence from the diffusion constant.

We end with more detailed comparison of resistivity with available theories. At our higher experimental temperatures we compare with FTLM, which is an exact technique, and find reasonable agreement (Fig.~\ref{fig:conductivity}). The experimental resistivity is systematically smaller than the FTLM calculation but within error bars. This may be a result of the uncertainty in determining $U/t$. At our lowest experimental temperatures, FTLM suffers from finite size effects which become relevant as correlation lengths approach the cluster size. For the $4 \times 4$ site cluster considered here, these effects limit FTLM resistivity calculations to $T/t \gtrsim 1$. 

Because our experiment explores low temperatures which are inaccessible to FTLM, we also compare with an approximate technique, single-site DMFT \cite{Kotliar2006} (Fig.~\ref{fig:conductivity}). We find the DMFT tends to overestimate the experimental resistivity at high temperatures. At our highest experimental temperatures, the DMFT resistivity is linear with a positive zero-temperature intercept. This linear scaling crosses over to a second linear scaling with a negative zero-temperature intercept around $T/t = 2$. This second linear region continues down to about $T/t = 0.8$ where the resistivity acquires a significant quadratic component. These regimes coincide with two different regimes observed in the DMFT compressibility \cite{supplement}. Previous DMFT studies at stronger interaction strengths have also observed these two linear regimes at intermediate temperatures, finding evidence for resilient quasiparticles in the lower temperature regime \cite{Xu2013, Deng2013}. We do not observe the change of slope in the resistivity expected near $T/t = 2$ in either the experimental data (within uncertainties) or the FTLM results. This suggests a need for comparison between more refined DMFT and exact theoretical approaches in the regime where this is possible.


Our experiment paves the way for future studies of the optical conductivity and thermopower, which can be examined near equilibrium using a similar approach. Both of these quantities might be expected to show anomalous scalings, as in the cuprates \cite{Varma1989, Hussey2008}. In line with theoretical work such as \cite{Xu2013, Deng2013}, searching for direct signatures of resilient quasiparticles using spectroscopic techniques \cite{Stewart2008} would also be very interesting. Further experimental studies will also provide important benchmarks for approximate theoretical methods, as the combination of low temperature, finite-doping, and dynamics is challenging for exact theoretical approaches.

\pagebreak

\section{Acknowledgements}

\begin{acknowledgments}
We thank Joseph Thywissen, Martin Zwierlein, and Sean Hartnoll for stimulating discussions. We thank Maxime Charlebois and Patrick S\'emon for contributions to the continuous time Monte Carlo impurity solver codes. We thank Dominic Bergeron for assistance with the analytic continuation and comparison of results with the Two-Particle-Self-Consistent (TPSC) approach. This work was supported by the NSF (grant no. DMR-1607277), the David and Lucile Packard Foundation (grant no. 2016-65128), the AFOSR Young Investigator Research Program (grant no. FA9550-16-1-0269), the Canada First Research Excellence Fund, the Natural Sciences and Engineering Research Council of Canada (NSERC) under grant RGPIN-2014-04584, the Research Chair in the Theory of Quantum Materials (AMST), and the Slovenian Research Agency Program P1-0044. Simulations were performed on computers provided by the Canadian Foundation for Innovation, the Minist\`ere de l'\'Education des Loisirs et du Sport (Qu\'ebec), Calcul Qu\'ebec, and Compute Canada.  W.S.B. was supported by an Alfred P. Sloan Foundation fellowship. P.T.B. was supported by the DoD through the NDSEG Fellowship Program. Data reported in this paper and the code required to reproduce the data analysis are archived on the Open Science Framework and GitHub \cite{Brown2018a, Kokalj2018, Hebert2018, Nourafkan2018}. 
\end{acknowledgments}


%

\pagebreak


\pagebreak
\clearpage
\setcounter{equation}{0}
\setcounter{figure}{0}

\renewcommand{\theparagraph}{\bf}
\renewcommand{\thefigure}{S\arabic{figure}}
\renewcommand{\theequation}{S\arabic{equation}}

\onecolumngrid
\flushbottom


\section{Methods}
We work with an equal spin mixture of $^6$Li hyperfine ground states $\ket{1}$ and $\ket{3}$, numbered up from the lowest energy state, which we label as spin up, $\ket{\uparrow}$, and spin down, $\ket{\downarrow}$, respectively. Our system is well described by the Fermi-Hubbard model with the Hamiltonian,

\begin{equation}
\mathcal{H} - \mu N = -t \sum_{\ensavg{\mathbf{i}, \mathbf{j}}, \sigma} \left(c^\dag_{\mathbf{i}, \sigma} c_{\mathbf{j}, \sigma} + \text{h.c.} \right) + U \sum_\mathbf{i}  n_{\mathbf{i},\uparrow} n_{\mathbf{i},\downarrow} - \mu \sum_{\mathbf{i},\sigma} n_{\mathbf{i},\sigma}, \label{eq:hubbard_h}
\end{equation}
where $c^\dag_{\mathbf{i}, \sigma}$ is the creation operator for a fermion on site $\mathbf{i} = (i_x, i_y)$ with spin $\sigma \in \{\uparrow, \downarrow\}$, $n_{\mathbf{i},\sigma} = c^\dag_{\mathbf{i},\sigma} c_{\mathbf{i},\sigma}$, $t$ is the hopping rate, $U$ is the on-site interaction, and $\mu$ is the chemical potential.

The experimental setup and basic parameters are described in detail in the supplement of ref.~\cite{Brown2017}. After preparing a 2D degenerate Fermi gas, we simultaneously load the optical lattice to a final depth of \SI{6.9(2)}{E_R} and the sinusoidally modulated potential with a \SI{50}{\milli\second} intensity ramp. We then turn off the sinusoidal modulation in approximately \SI{10}{\micro \second} using the spatial light modulator and observe the decay of the density pattern. 

We work at a field of \SI{616.0(2)}{G}. At this field and lattice depth, we find $U/t = 7.4(8)$ from a band structure calculation, which yields $t = \SI{925(10)}{\hertz}$, and spectroscopic measurement of $U=\SI{7.0(7)}{\kilo \hertz}$. 

To prepare clouds of variable temperature, we use two different protocols. To reach temperatures in the range $T/t = 0.3 - 2$, we hold the cloud in the final trapping configuration for variable time. The gas heats due to technical noise at a rate of $~3t$ per second. To reach even hotter temperatures, we modulate the lattice depth at a frequency of \SI{2}{\kilo \hertz}. To avoid losses, we perform this modulation at \SI{595}{G} where the interaction is weaker. Finally, we ramp the field to its final value, turn on the DMD potential, and follow the same protocol as before.

For low temperatures, our lattice provides all of the radial confinement. For temperatures hotter than $T/t \approx 3$, the compressibility of the gas is reduced and we must provide extra confinement to reach appropriate filling. Therefore we increase our trapping frequency using a \SI{1064}{\nano \meter} beam. 

To image the density of a single spin state, we freeze the motion of the lattice by ramping the lattice depth to $60 \ E_R$ in approximately \SI{100}{\micro \second}. We checked that this ramp effectively freezes the atomic motion by comparing the measured amplitude modulation without turning off the DMD potential and with turning off the DMD potential and then immediately ramping the lattice depth for our shortest wavelength at our lowest temperatures (where the modulation decays fastest). The modulation depths agreed, indicating that the atomic motion is effectively frozen well before the lattice reaches $60 \ E_R$.

\section{DMD calibration}
We engineer our deconfining and sinusoidally modulated potential using up to \SI{15}{\milli\watt} of \SI{650}{\nano\meter} coherent light derived from a tapered amplifier fed by a diode laser. Our spatial light modulator is a DLP Discovery 4100 with a DLP7000 digital micromirror device (DMD) in an imaging plane configuration. We image this light onto our atoms using two stages of demagnification. First, we demagnify the DMD image by a factor of 5, then we combine the DMD projection path with our imaging path on a dichroic mirror. Our imaging system demagnifies the light by an additional factor of 30. A single DMD micromirror has a pitch of \SI{13.68}{\micro\meter}, so approximately $8 \times 8$ mirrors determine the potential at a single lattice site. Our imaging system spatially filters the binary image, resulting in a smooth potential at the atoms.

Before each experiment, we load a series of two images into the DMD memory. The first is the sum of a deconfining Gaussian potential for flattening the atomic density in the central part of the cloud with a sinusoidal modulation pattern. The second is only the deconfining potential. The DMD displays these images successively after receiving a trigger. We use the ALP-4.2 API ``uninterruptible binary mode'' to keep the image on the DMD until the next trigger. The DMD transitions between images in approximately \SI{10}{\micro\second}. During this time, all mirrors go to the off state, and then the mirrors needed for the next image are turned to the on state. The motion of the mirrors is underdamped, and we observe the mirrors bouncing by measuring diffracted light on a photodiode.

We produce binary images from continuous potential profiles using the Floyd-Steinberg error diffusion algorithm \cite{Floyd1976}.

\section{Hydrodynamic model}
Hydrodynamics applies at long wavelengths and low frequencies when there are few conserved quantities, typically only mass, momentum, and energy. In most real materials, electrons cannot be treated hydrodynamically because of couplings to phonons and lattice defects which can absorb energy or momentum. For strongly interacting systems with no external couplings, hydrodynamics is applicable. In lattice systems the momentum is not conserved due to umklapp scattering, and only energy and particle number are conserved. In systems with weak umklapp (momentum relaxation) rate, we can also write down a ``momentum conservation'' equation including this relaxation rate. For a detailed discussion of when strongly interacting systems can be treated hydrodynamically, see \cite{Hartnoll2018}.
 
The simplest hydrodynamic theory we can write down accounts for conservation of mass, weak relaxation of momentum, and assumes that energy is decoupled from these two. The two equations describing this are
 \begin{eqnarray}
 \label{eq:continuity}
  \partial_t n(\mathbf{r}, t) &=& -\nabla \cdot \mathbf{J}(\mathbf{r}, t) \\
  \label{eq:modified_fick_law}
 \partial_t \mathbf{J}(\mathbf{r}, t) &=& - \Gamma \left( D \nabla n(\mathbf{r}, t) + \mathbf{J}(\mathbf{r}, t) \right),
 \end{eqnarray}
 where the first equation is the continuity equation, and the second equation reduces to Fick's law, $\mathbf{J}(\mathbf{r}, t) = - D \nabla n(\mathbf{r}, t)$, when $\partial_t \mathbf{J} = 0$. In the limit of strong momentum relaxation, a density modulation instantly creates a current satisfying Fick's law. However, for a finite relaxation rate $\Gamma$, the current does not instantly follow the changes in the density, and its time lag is described by eq.~\ref{eq:modified_fick_law}. We can alternatively understand eq.~\ref{eq:modified_fick_law} as a momentum ``conservation'' equation analogous to the Navier-Stokes equation including a weak momentum relaxation rate $\Gamma$, zero viscosity, and neglecting terms of higher order in linear response (e.g. terms proportional to the velocity squared).
 
 We spatially Fourier transform these equations and eliminate $J$ to find,
 \begin{equation}
 \partial_t^2 n_k + \Gamma \partial_t n_k + \Gamma Dk^2 n_k = 0, \label{eq:hydro_diffeq}
 \end{equation}
 which is the equation that appears in the main text. 
 
 This is the same differential equation that describes a damped harmonic oscillator provided we identify the natural frequency $\omega_o = \sqrt{D\Gamma} k$ and the damping rate as $\gamma = \Gamma /2$. This oscillator model crosses over from an underdamped to an overdamped regime at a modulation wavelength $\lambda = 4\pi \sqrt{D/\Gamma}$. In a system that can be described using quasiparticles, $\sqrt{D/\Gamma}$ is proportional to the mean free path.
 
 In the overdamped limit, $\Gamma \gg \sqrt{\Gamma D} k$, we recover diffusive behavior $n_k(t) \propto e^{-Dk^2t}$ for finite $k$. In the underdamped limit, we have sound-waves whose amplitude decays at rate $\Gamma/2$. If we take the limit $k = 0$, the current decays exponentially at rate $\Gamma$, identifying this as the momentum relaxation rate. The sound-wave and current relaxation rates differ because the sound wave carries both kinetic and potential energy, shared equally, whereas the uniform ($k=0$) current excitation carries only kinetic energy. As only the kinetic energy is damped, the sound-wave loses energy at half of the rate of the uniform current excitation.

 \section{Details of fitting the hydrodynamic model}
 
 For each temperature, we take images at a series of wavevectors $k_i$, $i=1, ..., N$ at times $t_{ij}$, $j=1, ..., j_i$. For each $(k_i, t_{ij})$ we determine the average 2D density profile. We produce a 1D representation of the modulation by averaging density in the central region of the cloud along the direction perpendicular to the modulation and fit the result to a sinusoidal pattern. 
 For a given wavevector, we first fit the shortest time, $t_{i1}=0$, modulation pattern with the sinusoidal amplitude, period, phase, and offset as free parameters. For later times, we fix the phase and the period, leaving only the amplitude and the offset as free parameters. From this procedure we extract a series of amplitudes, $a(k_i, t_{ij})$ with uncertainties $\sigma_{ij}$. Because the phase is fixed by the first pattern, a negative amplitude $a(k_i, t_{ij}) < 0$ indicates that the modulation pattern at time $t_{ij}$ is $180^\circ$ out of phase with the initial pattern.

 To compare our measurements with the hydrodynamic model, we write the solutions to eq.~\ref{eq:hydro_diffeq} which satisfy the boundary condition $\dot n(t=0) = 0$,
 \begin{equation}
 n(\Gamma, D, A, k, t) =  \frac{A}{2}\left(e^{\tilde \omega t} + e^{-\tilde \omega t} \right)e^{-\Gamma t /2}, \label{eq:hydro_diffeq_solns}
 \end{equation}
 where $\tilde \omega = \sqrt{\frac{\Gamma^2}{4} - \Gamma D k^2}$, and $A$ is the amplitude. In the underdamped limit eq.~\ref{eq:hydro_diffeq_solns} gives a damped cosine. In the overdamped limit it yields a product of a hyperbolic cosine factor and an exponential factor.
 
 Finally, we perform a non-linear least squares fit which minimizes
 \begin{equation}
 \sum_{i=1}^N \sum_{j=1}^{j_i} \frac{\left|a(k_i, t_{ij}) -  n(\Gamma, D, A_i, k_i, t_{ij}) \right|^2}{\sigma_{ij}^2},
 \end{equation}
 with the free parameters $\Gamma$, $D$, and $A_i$, $i=1, ..., N$. We determine the uncertainty in the fit parameters with a bootstrapping technique.
 
\section{Linear response theory}

To connect our hydrodynamic model for the density response of our Fermi-Hubbard system to quantities which can be calculated in theory, we consider the effect of perturbing our system with a time and spatial dependent potential, $v_\mathbf{i}(t) = F(t) \sin(\mathbf{k} \cdot \mathbf{r}_\mathbf{i})$. In this experiment we suppose $F$ is turned on slowly starting at $t = -\infty$ and switched off suddenly at $t = 0$, leading to $F(t) = e^{\eta t} \theta(-t)$, where $\eta$ parametrizes the slow turn on. If we suppose that $H_o$ is the Fermi-Hubbard Hamiltonian in the absence of this, perturbation, then the full Hamiltonian is $H(t) = H_o + H'(t)$. We can write the perturbation term as
\begin{eqnarray}
H'(t) &=& -\sum_\mathbf{i} v_\mathbf{i}(t) n_\mathbf{i} \\
&=& - \sum_\mathbf{k} v_{\mathbf{k}}(t) n_{-\mathbf{k}}, \label{eq:h_conj_field}
\end{eqnarray}
where $\mathbf{r}_\mathbf{i}$ is the position of site $\mathbf{i}$ and $n_k$ is the spatial Fourier transform of the density.

In linear response theory we think of $v$ as the force which is conjugate to the density response. Given a Hamiltonian of the form in eq.~\ref{eq:h_conj_field}, we can write the density reaction in terms of a response function $\Phi$
\begin{eqnarray}
\ensavg{\delta n_\mathbf{k}(t)} &=& \int_{-\infty}^t dt' \ \Phi(\mathbf{k}, t - t') v_\mathbf{k}(t') \label{eq:nk-resp}\\
\Phi(\mathbf{k}, t - t') &=& -\frac{i}{\hbar} \Theta(t - t') \ensavg{[n_\mathbf{k}(t), n_{-\mathbf{k}}(t')]},
\end{eqnarray}
where we used translational invariance of the unperturbed system, which ensures that only $v_\mathbf{k}$ contributes to the density response at $\mathbf{k}$. This equation says that the response of the density to the applied field is given by the density correlations, encapsulated in the retarded Green's function $\Phi$.

Fourier transforming eq.~\ref{eq:nk-resp} in time and space leads to,
\begin{eqnarray}
\ensavg{\delta n_\mathbf{k}(\omega)} &=& \chi(\mathbf{k}, \omega) v_\mathbf{k} (\omega)  \label{eq:nk-resp-frq}\\
\chi(\mathbf{k}, \omega) &=& \int_{-\infty}^\infty dt' \ e^{i\omega t'} \Phi(\mathbf{k}, t') \\
&=& -\frac{i}{\hbar} \int_0^\infty dt' \ e^{i(\omega + i\eta) t'} \Theta(t') \ensavg{\left[n_\mathbf{k}(t'), n_{-\mathbf{k}}(0) \right]} \label{eq:chi},
\end{eqnarray}
where we added an small positive imaginary part, $\eta$, to the frequency so that the integral converges. We refer to $\chi(\mathbf{k}, \omega)$ as the density response function. $\chi$ is analytic in the upper-half plane, its real part is symmetric in $\omega$, and its imaginary part is antisymmetric in $\omega$. 

\section{Nernst-Einstein equation}

In our experiment we have direct access to $\chi(\mathbf{k}, \omega)$ because we control the potential which is the generalized force that couples to the density. We want to measure the charge conductivity, $\sigma$, which is the response function for the current. The current is conjugate to the vector potential, which we do not control in this experiment. Fortunately, the continuity equation written using linear response relations connects the density response function with the conductivity,
\begin{equation}
\sigma'(\omega) = \lim_{k\rightarrow0} \frac{\omega}{k^2} \chi''(\mathbf{k},\omega),
\label{eq:einstein-lr}
\end{equation}
where $\sigma'(\omega)$ is the real part of the conductivity and $\chi''(\mathbf{k}, \omega)$ is the imaginary part of the density response function.

To apply this expression, we need to know the form of $\chi$. This is provided by our hydrodynamic model. Adding a force term to eq.~\ref{eq:hydro_diffeq} and using the definition of the response function from eq.~\ref{eq:nk-resp-frq} leads to the expression
\begin{equation}
\chi(\mathbf{k}, \omega) = \frac{\chi_c}{1 - \frac{i \omega}{k^2 D} - \frac{\omega^2}{k^2 D \Gamma}},
\label{eq:chi-hydro}
\end{equation}
where $\chi_c$ is the charge compressibility, $D$ is the diffusion constant, and $\Gamma$ is the momentum relaxation rate \cite{Kadanoff1963}. Inserting this expression for $\chi$ in eq.~\ref{eq:einstein-lr}, we find that the optical conductivity has a Lorentzian profile with half-width half-maximum $\Gamma$,
\begin{equation}
\sigma'(\omega) = \frac{\chi_c D}{1+\left(\frac{\omega}{\Gamma} \right)^2},
\label{eq:optical_cond_hydro_model}
\end{equation}
and the Nernst-Einstein relationship holds for the DC conductivity,

\begin{equation}
\sigma = \chi_c D.
\label{eq:einstein}
\end{equation}

The Nernst-Einstein relation is a consequence of eq.~\ref{eq:einstein-lr} which holds if the density has a diffusive mode at long times and large wave vectors. The Nernst-Einstein relation does not require the exact form for $\chi$ in eq.~\ref{eq:chi-hydro}. We can, for example, add $k^2$-dependence to $D$ or $\Gamma$.
 
\section{Linearity}

\begin{figure}[b]
\centering
\includegraphics[width=\textwidth]{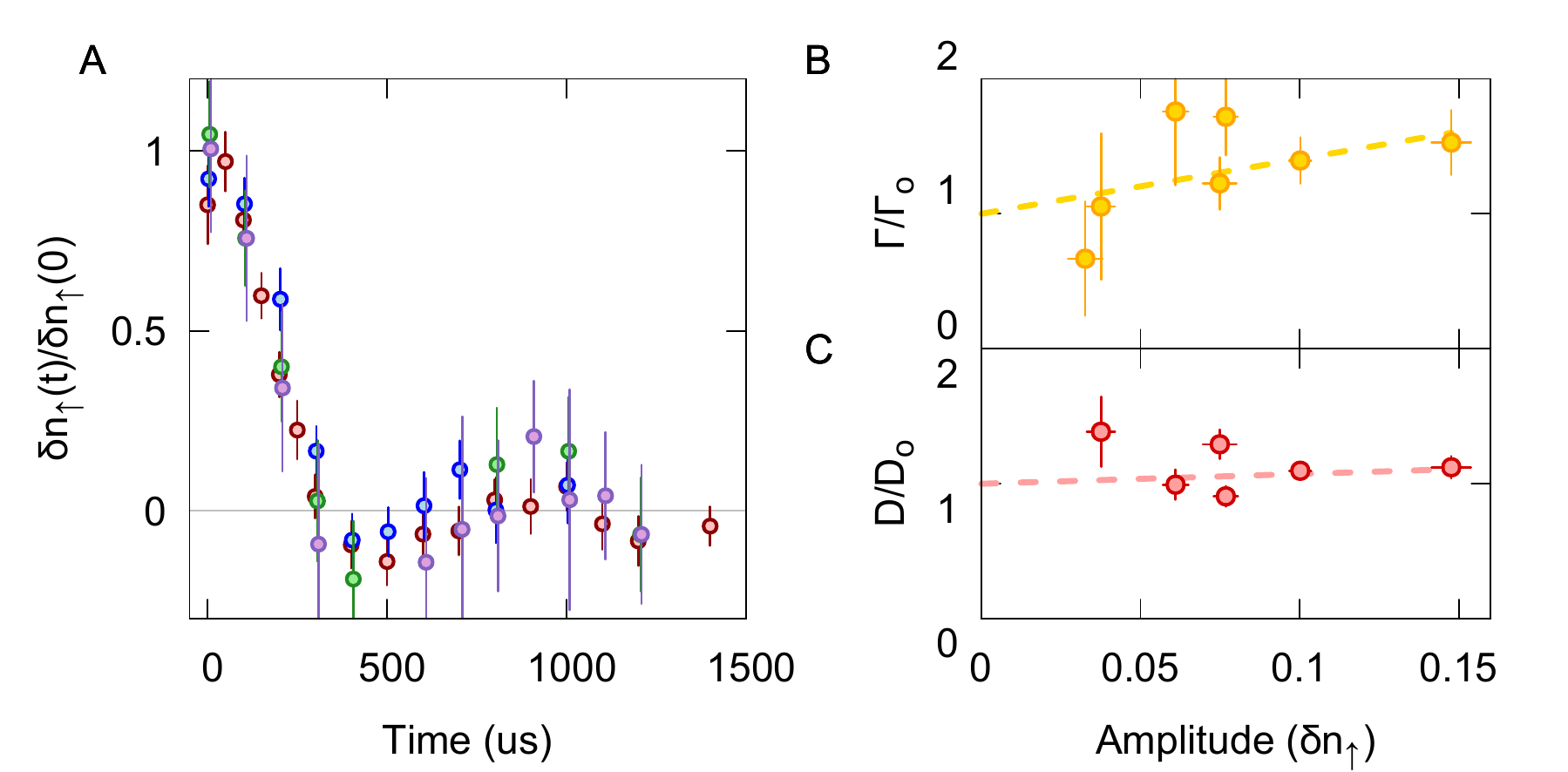}
\caption{{\bfseries Linearity of the density response.} \label{fig:linearity} ({\bfseries A}) Modulation amplitude versus decay time curves for selected initial amplitudes, $\delta n_\uparrow(t = 0) = 0.12$ (red), $0.08$ (blue), $0.055$ (green), and $0.035$ purple. We see a collapse after scaling the curves to the initial modulation amplitude, $\delta n_\uparrow (0)$, obtained from a fit. Each point is the average of $\approx 30$ images. ({\bfseries B}) Variation in fit parameter $\Gamma$ versus amplitude for the curves shown in a (red) and a linear fit to the results (dashed line). $\Gamma$ is normalized by the extrapolated zero-amplitude value, $\Gamma_o$. ({\bfseries C}) Variation in fit parameter $D$ versus amplitude for the curves shown in a (red) and a linear fit to these results (dashed line). $D$ is normalized by the extrapolated zero-amplitude value, $D_o$. Error bars sem.}
\end{figure}

To assess the possibility of non-linear effects which are not included in our hydrodynamic model, we varied the initial amplitude of the density modulation at a fixed wavevector. For each curve, we fit a value for $\Gamma$ and $D$, to test how the fitted model parameters change with amplitude. The amplitude versus time curves are shown in Fig.~\ref{fig:linearity}A for $\lambda \approx 12$ sites and temperature $T/t = 0.4(1)$. For each initial amplitude, we fit values for $\Gamma$ and $D$ using our hydrodynamic model. The fit results are shown in Fig.~\ref{fig:linearity}B,C. We find that the apparent $\Gamma$ increases with increasing amplitude, and the apparent $D$ is weakly effected by increasing amplitude. To establish an upper bound on the size of this effect, we perform a linear fit to the hydrodynamic parameters versus amplitude and extrapolate a `zero-amplitude' value. We normalize the curve fit parameters by these values in Fig.~\ref{fig:linearity}B,C. Based on the statistical error in our fit lines, we find that at a typical experimental amplitude of $\delta n_\uparrow = 0.07$, $\Gamma$ is increased by a factor of $1.4(4)$ and $D$ by a factor of $1.06(10)$. Our extracted values for $\Gamma$ appear to increase with amplitude, but the statistical error bar is quite large. This is due to the weak dependence of our model on the value of $\Gamma$. In the main text, we are able to obtain smaller error bars on $\Gamma$ by simultaneously fitting decay curves at different modulation wavelengths. That approach is not feasible here because the degree of linearity may depend on modulation wavelength.

A related but distinct type of non-linearity is dependence of the charge compressibility on density. As the total density approaches half-filling, the compressibility decreases. Therefore, the chemical potential modulation we apply tends to decrease the density at the minimum chemical potential values more than it increases the density at the maximum chemical potential values. This can lead to the density modulation deviating from a sine wave. At $T/t = 0.4$, the compressibility decreases by $\approx 20~\%$ between $\ensavg{n} = 0.8 - 0.9$. At higher temperature, $T/t = 4$, the compressibility decreases by $\approx 2~\%$. We resolve this effect as a shift in the average density between the initial density modulation pattern and the long-time equilibrium density. This effect is largest at the coldest temperatures, and is at most $\delta n_\uparrow \approx 0.03$, which is comparable to the uncertainty in our density.

\section{Temperature fitting}


\begin{figure}[b]
\centering
\includegraphics[width=0.3\textwidth]{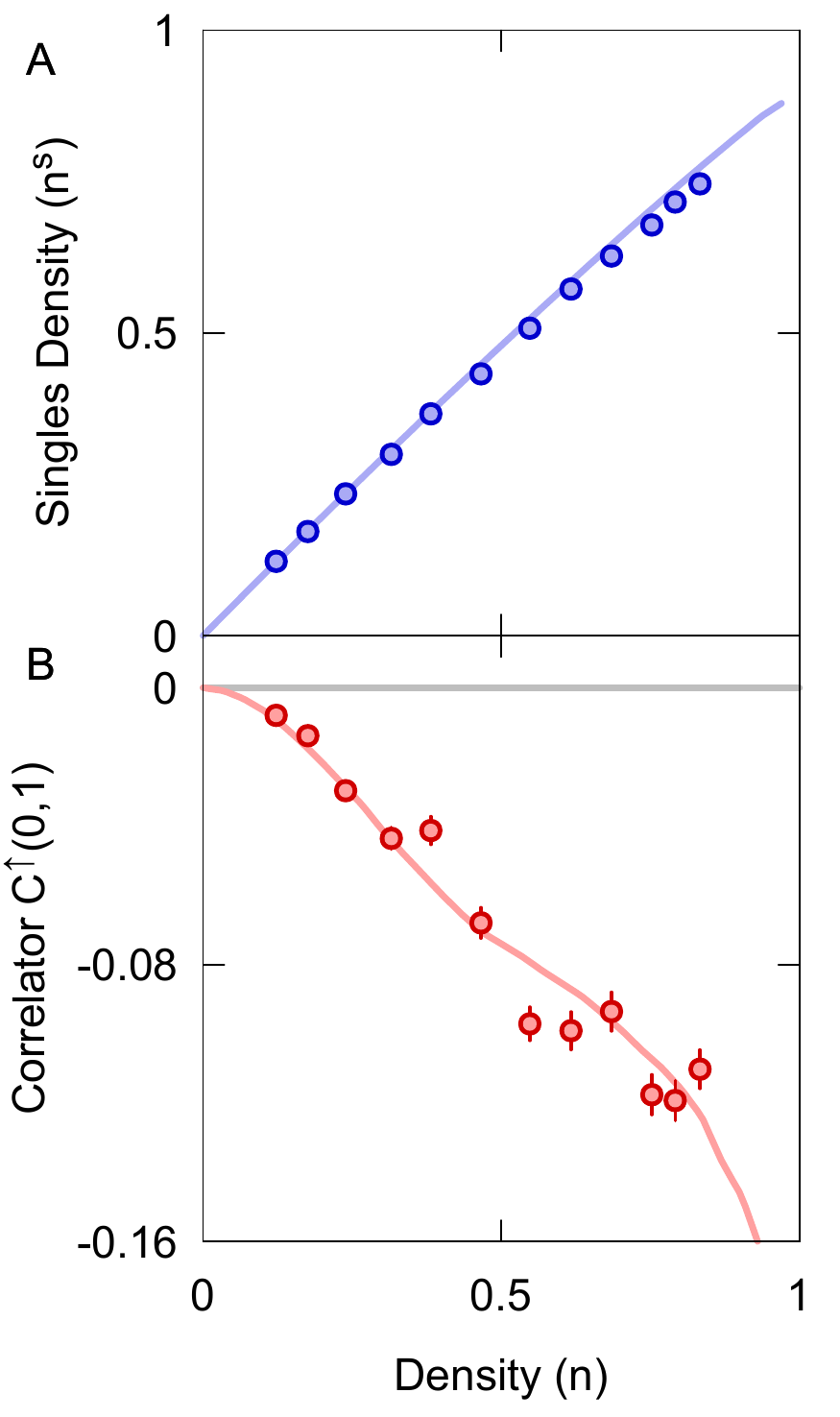}
\caption{{\bfseries Temperature fitting.} \label{fig:fit_temp} ({\bfseries A}) Experimental density versus singles density (points) and DQMC results (line). The DQMC results have been corrected for the experimental detection efficiency. This fit yields $T/t = 0.5(1)$. ({\bfseries B}) Spin-up spin-up density correlator, $C^\uparrow(0,1)$, for nearest-neighbor sites. Error bars sem.}
\end{figure}

We determine the temperature of our clouds by fitting to determinantal quantum Monte Carlo (DQMC) results. Spin resolved density $\ensavg{n_\uparrow}, \ensavg{n_\downarrow}$, correlation functions $\ensavg{n_\uparrow(\mathbf{r})n_\uparrow(0)}_c$, and singles density $\ensavg{n^s}$ data is generated at $U/t = 8$ on a grid of chemical potentials and temperatures.  The total density and correlation functions are then interpolated on a regular grid of density and temperature points. 

We use these interpolating functions to simultaneously fit the singles density and the single-spin component correlations versus the total density. The only free parameter is the temperature. We apply an imaging fidelity correction of $f=0.97$ based on our measured hopping and loss rates during imaging. An example fit is shown in Fig.~\ref{fig:fit_temp}.

\section{Compressibility}

\begin{figure}
\centering
\includegraphics[width=\textwidth]{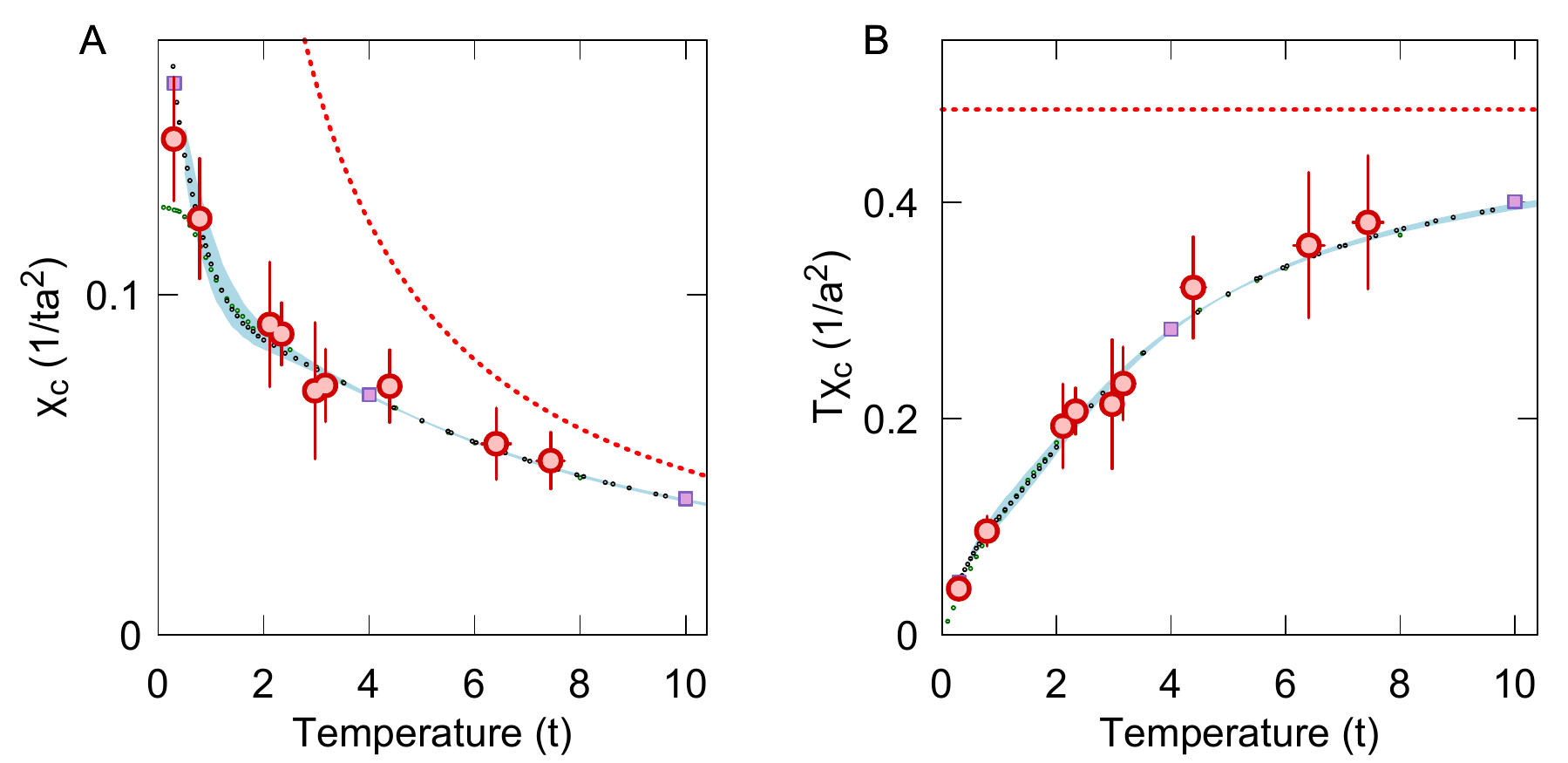}
\caption{{\bfseries Compressibility.} \label{fig:compressibility} ({\bfseries A}) Results for the charge compressibility, $\chi_c$. Experimental results (red points), FTLM (blue band), DQMC at $\ensavg{n} = 0.83$ and $U/t = 7.5$ (grey points), single-site DMFT at $\ensavg{n} = 0.825$ and $U/t = 7.5$ (green points), cellular DMFT (purple squares), and the high-temperature limit scaling $T \chi_c(T) = n(1-n/2)$ (red dashed line). ({\bfseries B}) Results for the charge compressibility times the temperature, $T \chi_c$, using the same color scheme as {\bfseries A}. Error bars sem.}
\end{figure}

We measure the compressibility of our gas in a harmonic trap with no additional potential provided by the DMD. For low temperatures, the lattice beams provide all the radial confinement, leading to $\bar \omega = (2\pi) 185(10) \text{ Hz}$. For hotter temperatures, we use a circular beam to provide extra confinement, leading to $\bar \omega = (2\pi) 280(10) \text{ Hz}$.  

We determine the harmonic trapping frequencies by fitting the density and nearest-neighbor density correlation profiles of a weakly interacting gas obtained at a field of \SI{568.0(1)}{G}, near the noninteracting point of the $\ket{1}-\ket{3}$ mixture, to the expected values for a non-interacting Fermi gas. These are determined from
\begin{eqnarray}
n_\uparrow &=& \frac{1}{N}\sum_\mathbf{k} f(\epsilon_k - \mu, T)\\
\frac{1}{4} C^\uparrow(\mathbf{d}) &=& - \frac{1}{N^2}\left\vert \sum_\mathbf{k} f(\epsilon_k - \mu,T)e^{-i\mathbf{k} \cdot \mathbf{d}} \right\vert^2,
\end{eqnarray}
where $f(\epsilon,T)$ is the Fermi function, and $\mathbf{k}$ runs over the $N$ allowed lattice momenta. Distance and energy scales are measured in units of the lattice constant and hopping respectively. We assume a harmonic trapping potential, $\mu(r) = \mu_o - \frac{1}{2} m \omega^2 r^2$ and fit our cloud profiles with $\mu_o$, $\omega$, and $T$ as free parameters.

After determining the trapping frequency, we compute the compressibility according to,
\begin{equation}
\left. \left(\frac{\partial n}{\partial \mu}\right) \right \vert_T = - \frac{1}{m\omega^2} \left(\frac{1}{r} \frac{\partial n}{\partial r} \right).
\end{equation}

Our cloud is slightly elliptic, with an aspect ratio of $\omega_x/\omega_y \approx 1.2$. Prior to determining the trapping frequency we perform an azimuthal average, which effectively rescales our coordinates $\left(x,y\right) \rightarrow \left( x \sqrt{\frac{\omega_x}{\omega_y}}, y \sqrt{\frac{\omega_y}{\omega_x}} \right)$. We measure $r$ in these coordinates above, therefore our fitting procedures yields $\bar \omega = \sqrt{\omega_x \omega_y}$.

We compare our measured compressibility with DQMC, FTLM, and DMFT in Fig.~\ref{fig:compressibility}A. The DQMC and FTLM compressibilities agree well with the experimental data and do not saturate at low temperatures. In contrast, the single-site DMFT compressibility saturates at $T/t \approx 1$. The increasing compressibility below this temperature may be associated with short-range correlations \cite{Kokalj2017}, which are not accounted for by single-site DMFT. Cellular DMFT results using a $2 \times 2$ plaquette gives excellent agreement with DQMC, supporting this interpretation.

At high temperatures, the compressibility is expected to scale as $1/T$ with $T \chi_c(T) = n(1-n/2)$, for finite $U$ \cite{Perepelitsky2016}. We plot the compressibility times the temperature, $T \chi_c(T)$ in Fig.~\ref{fig:compressibility}B. $T \chi_c$ has not yet saturated in the temperature range considered here. We expect that saturation occurs at temperatures much hotter than the bandwidth, $T \gg 8t$.

\section{Thermoelectric effects}

We prepare our sample in thermal equilibrium, therefore there  are initially no thermal gradients. However, thermal gradients may be generated during the subsequent dynamics. To check this possibility, we looked at a wave vector with underdamped oscillations at our lowest temperature. We measured nearest-neighbor correlations, $C^\uparrow(0,1)$, at the time where the amplitude first crosses zero. Here the density is flat, and any spatial variations in the correlator must be due to thermal gradients. We did not find any evidence for generation of thermal gradients.

Thermoelectric coupling is primarily due to two effects. The first is thermodynamic, and is described by the thermoelectric susceptibility, $\zeta = -\frac{\partial^2 \Omega}{\partial \mu \partial T} = \left.\frac{\partial n}{\partial T} \right\vert_\mu = \left.\frac{\partial S}{\partial \mu} \right\vert_T$, where $\Omega = \epsilon - ST - n \mu$ is the grand potential and $S$ is the entropy. This is a static quantity, and can be computed, e.g., by FTLM. In the whole temperature regime accessible by FTLM we find $|\zeta| \lesssim 0.015 t^{-1}$. This is small in the sense that generating a density gradient of $0.01 a^{-1}$ requires a large temperature gradient of $\nabla T \approx 0.8 ta^{-1}$.

The Seebeck coefficient is more difficult to calculate. Using the Mott-Heikes approximation \cite{Palsson1998, Shastry2009} \, or the Kelvin formula \cite{Peterson2010,Arsenault2013,Kokalj2015} \, gives a small Seebeck coefficient due to a maximum of entropy which occurs close to $\ensavg{n} \approx 0.83$. This is in agreement with previous observations using different models or techniques \cite{Palsson1998, Jaklic2000}. A detailed description of particle diffusion in the presence of thermoelectric effects can be found in Ref.~\cite{Hartnoll2014}. In the bad-metallic or high-T regime where the Kelvin formula is a good approximation for the Seebeck coefficient \cite{Peterson2010, Kokalj2015}, the effect of thermoelectric coupling on particle diffusion is negligible.

\section{Comparison of theory techniques}
In the main text, we compare experimental results with three theory techniques, DQMC, FTLM, and DMFT. Each of these have different strengths and weaknesses which makes one or another more suitable for certain comparisons. We provide a broad outline their strengths and weaknesses in this section, and more detailed information on each technique in the following sections.

DQMC is the method of choice for calculating static quantities because it is an exact technique which can access the lowest temperatures we reach in the experiment. On the other hand, dynamical quantities are difficult to extract, as DQMC yields imaginary time Green's functions which must be analytically continued to real time. Analytic continuation is an ill-posed problem, and the statistical uncertainty of DQMC data further complicates matters. In the main text, we use DQMC to compute only static quantities, including the singles density, spin-up density correlator, and compressibility.

FTLM is an exact technique which provides direct access to both static and dynamic correlators but is not capable of reaching temperatures as low as DQMC. The minimum temperature it can access is limited by finite size effects. When correlation lengths exceed the size of the small system used, the results no longer reflect the behavior of the system in the thermodynamic limit. In the main text we use FTLM to compute all dynamic quantities which can be obtained from correlators (momentum relaxation rate and conductivity). We also provide the FTLM compressibility in the supplement to verify that this agrees with the DQMC result.

DMFT is an \emph{approximate} technique which maps an interacting problem onto a self-consistent quantum impurity problem. It becomes exact in infinite dimensions. The impurity problem can be solved using a variety of techniques. When exact diagonalization is used to solve the impurity problem, dynamical quantities can be obtained directly, without analytic continuation. DMFT calculations can be performed at lower temperatures than attainable by DQMC or DMFT. In the main text we use DMFT to compute all dynamical quantities which can be obtained from correlators. We also provide the DMFT compressibility in the supplement to verify that this agrees with the DQMC result. This comparison provides one test of the DMFT approximations.

\section{Determinantal quantum Monte Carlo (DQMC)}

We perform DQMC calculations using the Quantum Electron Simulation Toolbox (QUEST) \cite{Varney2009} \, on an $8 \times 8$ homogeneous square lattice. The inverse temperature is split into $L=40$ imaginary time slices, where $L \Delta \tau = \beta$. We perform 5000 warm up sweeps, 50000 measurement sweeps and between 100 and 1000 passes to accumulate adequate statistics. We find that the sign problem at $\ensavg{n} = 0.83$ and $U/t = 7.5$ becomes important below $T/t = 0.5$. Reliable results in the range $T/t = 0.3-0.5$ can be obtained with additional statistics. Below $T/t = 0.27$, the sign approaches zero.

\section{Finite-temperature Lanczos method (FTLM)}

The finite-temperature Lanczos method (FTLM) \cite{Jaklic2000, Kokalj2013} \, is an exact diagonalization approach on small clusters ($4\times 4$ in our case). The method employs Lanczos diagonalization, which yields exact extremal eigenstates and effective or approximate eigenstates in the middle of the many-body spectrum.  These states are together with the aid of the sampling over random initial vector used to calculate finite-$T$ properties. Results have unwelcome finite size effects, which are large below some temperature $T_\textrm{fs}$, but are small at $T>T_\textrm{fs}$ due to shorter correlation lengths at high $T$. To reduce the finite size effects we employ averaging over twisted boundary condition \cite{Poilblanc1991} \, (with $N_\Theta=64$ different boundary conditions) and summation over all symmetry sectors, e.g., we use the grand canonical ensemble.  We do not show FTLM results for $T<T_\textrm{fs}$.

Dynamical quantities like the optical conductivity $\sigma(\omega)$ are calculated as correlation functions via evaluation of matrix elements of, e.g., the current operator, between the effective many body eigenstates, which are in this case obtained from two separate Lanczos procedures \cite{Jaklic2000, Kokalj2017}. The spectra, represented as a sum of weighted delta functions $\delta(\omega-\omega_i)$, needs to be further smoothed or broadened by $\eta$. Due to the exponential number of many body states and a very dense spectra, particularly at elevated $T$, the broadening can be relatively small and in our case $\eta\sim0.1t \ll \Gamma$. FTLM gives very good results for the static quantities with negligible finite size dependence and no additional broadening needed. Dynamical quantities are more challenging and in particular the optical conductivity at finite frequencies is quite robust while at the lowest frequencies it still shows a small dependence on cluster size or shape and on the broadening used. We estimate the overall uncertainty for the most challenging dc conductivity to be at most 15 percent.

\begin{figure}
\centering
\includegraphics[width=0.85\textwidth]{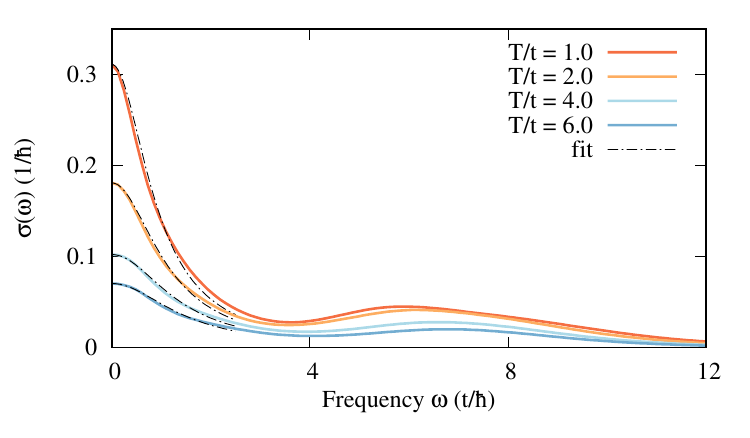}
\caption{{\bfseries Optical conductivity} \label{fig:optical_cond} 
Real part of the optical conductivity $\sigma(\omega)$ for several $T$, $U=7.5t$ and $n=0.83$ as calculated with the FTLM on a 16 site cluster. With dashed black line a Lorentzian fit in the low frequency regime ($0<\omega<2.5t/\hbar$) is shown.}
\end{figure}

The optical conductivity (Fig.~\ref{fig:optical_cond}) for our parameters exhibits a Drude peak at low $\omega$ and a separated Hubbard band at $\omega\sim U$. We fit the low-$\omega$ Drude peak to a Lorentzian, which is the form predicted by our hydrodynamic model (eq.~\ref{eq:optical_cond_hydro_model}), and extract the momentum relaxation rate $\Gamma$ presented in the main text.

\section{Dynamical mean-field theory (DMFT)}

We describe the dynamical mean-field theory self-consistency loop and then the various impurity solvers used in this paper.

\subsection{Self-consistency} Dynamical mean-field theory is exact in infinite dimension~\cite{Georges1992}. In finite dimension, it approximates the interacting problem by solving a self-consistent quantum impurity problem. A quantum impurity problem is the problem of a single site connected to an infinite bath of non-interacting electrons. The self-consistency is achieved by taking the same self-energy for both the quantum impurity problem and the lattice Green's function, and then requiring that the lattice Green's function projected on a single site equals the impurity Green's function. This takes into account both the localized physics of the atom and the itinerant character of the metal, in competition. When spatial correlations become important in low dimension, a cluster replaces the single site. The latter is known as Cellular Dynamical Mean-Field Theory (CDMFT)~\cite{Kotliar2001}.

We describe the CDMFT procedure. Single site DMFT is a special case where the cluster is replaced by a single site. For the Hubbard Hamiltonian, defined in eq.~\ref{eq:hubbard_h}, we write formally an effective action containing an hybridization function $\hat{\Delta}(\tau-\tau')$ that describes the degrees of freedom outside the cluster (the bath) as a time-dependent hopping within the cluster (which is easily pictured from a Feynman path integral point of view),
\begin{eqnarray}
S_{\mathrm{eff}}=\int_{0}^{\beta }d\tau d\tau ^{\prime }\Psi _{d}^{\dagger
}(\tau )\left[ (\partial /\partial \tau +\mu -\hat{t})\delta(\tau-\tau') - \hat{\Delta}(\tau-\tau') \right]\Psi _{d}(\tau ^{\prime
}) \nonumber\\
+ U\sum_{\mu }\int_{0}^{\beta }d\tau n_{\mu \uparrow }n_{\mu \downarrow }.
\end{eqnarray}%
Hats denote matrices in the cluster degrees of freedom labeled by the Greek letters $%
\mu ,\nu $. Here, $\hat{t}$ is the hopping of the original Hamiltonian within the cluster. For the case of a $2\times 2$ plaquette, the spinor is defined by
$\Psi _{d}^{\dagger }\equiv (d_{1\uparrow }^{\dagger },\dots ,d_{4\uparrow
}^{\dagger },d_{1\downarrow }^{\dagger },\dots ,d_{4\downarrow
}^{\dagger })$.  Physically, this
action corresponds to a cluster embedded in a self-consistently determined medium.

Given the effective action with a starting guess for $\hat{\Delta}(\tau-\tau')$, the cluster propagator $\widehat{G}_{c}$ is solved with three methods: Two variants of continuous-time quantum Monte Carlo, and exact diagonalization, on which we comment further below.  Once the cluster Green's function is obtained, we extract the cluster self
energy from $\hat{\Sigma}_{c}=\hat{\mathcal{G}}_{0}^{-1}-\widehat{G}%
_{c}^{-1} $ where $\mathcal{G}_{0}^{-1}$ is the quantity in square brackets in the quadratic part of the action, while $G_{\mu \nu ,\sigma }\equiv -\langle Td_{\mu
	\sigma }(\tau )d_{\nu \sigma }^{\dagger }(0)\rangle $ is the imaginary-time-ordered Green's function. Using the self-consistency condition in Matsubara frequency,
\begin{equation}
i\omega _{n} +\mu -\hat{t} - \hat{\Delta}(i\omega _{n})=\left[ \frac{N_{c}}{(2\pi )^{2}}%
\int d\tilde{\mathbf{k}}\;\widehat{G}(\widetilde{\mathbf{k}},i\omega _{n})\right]
^{-1}+\hat{\Sigma}_{c}(i\omega _{n})  \label{selfcon}
\end{equation}%
with
\begin{equation}
\widehat{G}(\widetilde{\mathbf{k}},i\omega _{n})=\left[ i\omega _{n}+\mu -\hat{t}(%
\widetilde{\mathbf{k}})-{\hat{\Sigma}_{c}}(i\omega _{n})\right] ^{-1},
\label{Lattice_G}
\end{equation}%
we recompute the hybridization function $\hat{\Delta}(i\omega _{n})$ and iterate till
convergence. Here $\hat{t}(\widetilde{\mathbf{k}})$ is the Fourier transform of the
superlattice hopping matrix, $N_c$ is the number of sites within the cluster and the integral over $\widetilde{\mathbf{k}}$ is performed over
the reduced Brillouin zone of the superlattice.

\subsection{Impurity solvers} The continuous-time quantum Monte Carlo solvers sample observables with a Markov chain defined in the space of Feynman diagrams of all orders. In the continuous-time auxiliary field method (CT-AUX)~\cite{Gull2008}, the action is expanded in powers of the Hubbard interaction. This approach works better when $U$ is less than the bandwidth. Expansion in powers of the hybridization function generates the so-called CT-HYB solver~\cite{Werner2006,Werner2006a,Haule2007a}, which works better at values of $U$ larger than the bandwidth~\cite{Gull2007}. For CT-HYB, we use a program that contains several improvements for speed~\cite{Semon2014}. The results are obtained from an average over the last 20 converged iterations and typically between $10^8$ and $3\times 10^9$ Monte Carlo updates. High frequency tails are usually of higher quality in the CT-AUX approach.

In the exact-diagonalization approach~\cite{Caffarel1994} \, that we used for single-site DMFT, the impurity problem is represented by an Anderson-like Hamiltonian $H_{\mathrm{imp}}$ with a
discrete number of bath orbitals (here 5 for each spin component) coupled to the impurity 
\begin{eqnarray}
H_{\mathrm{imp}} \equiv \sum_{m\sigma }\epsilon _{m\sigma
}a_{m\sigma }^{\dagger}a_{m\sigma } +\sum_{m \sigma }(V_{m \sigma }a_{m\sigma }^{\dagger}c_{\sigma }+\mathrm{h.c.})+Un_{\uparrow }n_{\downarrow }.
\end{eqnarray}%
Here, $m =1...5$ for each spin component such that we have $10$ bath energy
levels $\epsilon _{m\sigma }$ coupled to the impurity via the
bath-cluster hybridization matrix $V_{m \sigma }$. The hybridization function is obtained from
\begin{equation}
	\Delta_\sigma(i\omega_n)=\sum_{m}\frac{V_{m\sigma}^2}{i\omega_n-\epsilon_{m\sigma}}.
\end{equation}
The parameters $\epsilon_{m \sigma}$ and $V_{m \sigma}$ 
are determined by imposing the self-consistency condition in Eq.~\ref%
{selfcon} using a conjugate gradient minimization algorithm with a distance
function
\begin{equation}
d=\frac{1}{N_{max}}\sum_{n=0}^{N_{max}}\left\vert \left( \hat{\mathcal{G}}%
_{0}^{\prime -1}(i\omega _{n})-\hat{\mathcal{G}}_{0}^{-1}(i\omega
_{n})\right) \right\vert ^{2}  \label{distance}
\end{equation}%
where $N_{max}$ is the largest Matsubara frequency index, determined by choosing a high-energy cutoff of about $2000$ (energies are given in units of hopping $t,$ and we take $\hbar =1$ and $k_{B}=1$). The distance function in Eq.(\ref{distance})
is computed on the imaginary frequency axis since the hybridization function is a smooth function on that axis. We take a convergence criterion of $10^{-5}$ for the distance. We checked that the compressibility agrees with the continuous-time solvers to three significant digits. 

\subsection{Conductivity calculation}

Since in single-site DMFT the vertex corrections vanish, the optical conductivity is calculated from the single-particle spectral weight using
\begin{eqnarray}
\sigma'(\omega)=\frac{\chi^{\prime\prime}_{jj}(\omega)}{\omega} & = & \pi  \sum_{\sigma} \int_{-4t}^{4t} \! \mathrm{d}\varepsilon \int \! \mathrm{d}\omega'\, \mathcal{T}(\varepsilon)\, \mathcal{A}(\varepsilon, \omega')\, \mathcal{A}(\varepsilon, \omega' + \omega) \nonumber \\
& & \times \frac{\left[ f(\omega') - f(\omega' + \omega) \right]}{\omega}  ,
\label{Analytic_expression_chi}
\end{eqnarray}
where $f$ is the Fermi-Dirac distribution, $\chi^{\prime\prime}_{jj}(\omega)$ is the imaginary part of the current response function, and $\mathcal{A}(\mathbf{k},\omega)$ is the spectral function containing the non-interacting square-lattice dispersion $\varepsilon_{\mathbf{k}}$ and the impurity self-energy, and normalized so that $\int \! \mathrm{d}\omega\, \mathcal{A}(\mathbf{k},\omega) = 1$. Here, the usual integral over wave-vectors has been replaced by an integral over the band energies $\varepsilon$ weighted by the longitudinal transport function \cite{Arsenault2013a} 
\begin{equation}
\mathcal{T}(\varepsilon) = \sum_{\mathbf{k}} \left( \frac{\partial \varepsilon_{\mathbf{k}}}{\partial k_x} \right)^2 \delta(\varepsilon - \varepsilon_{\mathbf{k}}) = -\frac{1}{2}\int_{-4t}^{\varepsilon} \! z N_0(z) \, \mathrm{d}z
\end{equation}  
containing the non-interacting density of states $N_0$, normalized so that $\int \! N_0(z)\, \mathrm{d}z = 1$.

The real part of the conductivity obeys the $f$-sum rule in the following form~\cite{Maldague1977}
\begin{equation}
\int \frac{d\omega}{\pi}\sigma'(\omega)=\frac{1}{N}\sum_{\mathbf{k\sigma}}\frac{\partial^2\epsilon_\mathbf{k}}{\partial k_x^2}\left< n_\mathbf{k\sigma}\right> = -\frac{1}{2}E_{kinetic},
\end{equation}
where $\left< n_\mathbf{k\sigma}\right>$ is the expectation value of the occupation number in state $\mathbf{k}\sigma$ and $E_{kinetic}$ is the expectation value of the kinetic energy for this two-dimensional system with nearest-neighbor hopping only. This means that even in situations where the optical conductivity is dominated by a Drude peak whose width is temperature independent, as in the range $4<T<8$ in the inset of Fig. 3, the DC conductivity can decrease with temperature because at high temperature the kinetic energy decreases as $1/T$~\cite{Calandra2003}
. 

The exact diagonalization method, which is used to obtain the momentum relaxation rate and conductivity presented in the main text, allows one to obtain results directly on the real axis. However, the discrete nature of the bath introduces some uncertainty because the discrete energy levels must be broadened as Lorentzians of width $\eta$. When there is a range of $\eta$ where the results are independent of $\eta$, one can be confident of the results. The error bars on the DMFT results in the main text correspond to the difference between $\eta=0.1$ and a five times smaller value, $\eta=0.02$. The estimate of of the momentum relaxation rate $\Gamma$ is much less sensitive to $\eta$ than the DC conductivity. 

We checked that the resistivity obtained in single-site DMFT is independent of the solver used by comparing the exact diagonalization solver result with Pad\'e analytic continuation of the CT-AUX solver. We found good agreement up to $T/t=2$. At higher temperature we cannot make this comparison because analytic continuation is unreliable. We also verified that $\Gamma$ agrees for the two DMFT solvers.


\begin{thebibliography}{60}%
\makeatletter
\providecommand \@ifxundefined [1]{%
 \@ifx{#1\undefined}
}%
\providecommand \@ifnum [1]{%
 \ifnum #1\expandafter \@firstoftwo
 \else \expandafter \@secondoftwo
 \fi
}%
\providecommand \@ifx [1]{%
 \ifx #1\expandafter \@firstoftwo
 \else \expandafter \@secondoftwo
 \fi
}%
\providecommand \natexlab [1]{#1}%
\providecommand \enquote  [1]{``#1''}%
\providecommand \bibnamefont  [1]{#1}%
\providecommand \bibfnamefont [1]{#1}%
\providecommand \citenamefont [1]{#1}%
\providecommand \href@noop [0]{\@secondoftwo}%
\providecommand \href [0]{\begingroup \@sanitize@url \@href}%
\providecommand \@href[1]{\@@startlink{#1}\@@href}%
\providecommand \@@href[1]{\endgroup#1\@@endlink}%
\providecommand \@sanitize@url [0]{\catcode `\\12\catcode `\$12\catcode
  `\&12\catcode `\#12\catcode `\^12\catcode `\_12\catcode `\%12\relax}%
\providecommand \@@startlink[1]{}%
\providecommand \@@endlink[0]{}%
\providecommand \url  [0]{\begingroup\@sanitize@url \@url }%
\providecommand \@url [1]{\endgroup\@href {#1}{\urlprefix }}%
\providecommand \urlprefix  [0]{URL }%
\providecommand \Eprint [0]{\href }%
\providecommand \doibase [0]{http://dx.doi.org/}%
\providecommand \selectlanguage [0]{\@gobble}%
\providecommand \bibinfo  [0]{\@secondoftwo}%
\providecommand \bibfield  [0]{\@secondoftwo}%
\providecommand \translation [1]{[#1]}%
\providecommand \BibitemOpen [0]{}%
\providecommand \bibitemStop [0]{}%
\providecommand \bibitemNoStop [0]{.\EOS\space}%
\providecommand \EOS [0]{\spacefactor3000\relax}%
\providecommand \BibitemShut  [1]{\csname bibitem#1\endcsname}%
\let\auto@bib@innerbib\@empty
\bibitem [{\citenamefont {Coleman}(2015)}]{Coleman2015}%
  \BibitemOpen
  \bibfield  {author} {\bibinfo {author} {\bibfnamefont {P.}~\bibnamefont
  {Coleman}},\ }\href {\doibase 10.1017/CBO9781139020916} {\emph {\bibinfo
  {title} {Introduction to Many-Body Physics}}}\ (\bibinfo  {publisher}
  {Cambridge University Press},\ \bibinfo {year} {2015})\BibitemShut {NoStop}%
\bibitem [{\citenamefont {Ioffe}\ and\ \citenamefont
  {Regel}(1960)}]{Ioffe1960}%
  \BibitemOpen
  \bibfield  {author} {\bibinfo {author} {\bibfnamefont {A.}~\bibnamefont
  {Ioffe}}\ and\ \bibinfo {author} {\bibfnamefont {A.}~\bibnamefont {Regel}},\
  }\href {https://elibrary.ru/item.asp?id=21765390} {\bibfield  {journal}
  {\bibinfo  {journal} {Prog. Semicond.}\ }\textbf {\bibinfo {volume} {4}},\
  \bibinfo {pages} {237} (\bibinfo {year} {1960})}\BibitemShut {NoStop}%
\bibitem [{\citenamefont {Mott}(1972)}]{Mott1972}%
  \BibitemOpen
  \bibfield  {author} {\bibinfo {author} {\bibfnamefont {N.~F.}\ \bibnamefont
  {Mott}},\ }\href {\doibase 10.1080/14786437208226973} {\bibfield  {journal}
  {\bibinfo  {journal} {Philos. Mag.}\ }\textbf {\bibinfo {volume} {26}},\
  \bibinfo {pages} {1015} (\bibinfo {year} {1972})}\BibitemShut {NoStop}%
\bibitem [{\citenamefont {Hussey}(2008)}]{Hussey2008}%
  \BibitemOpen
  \bibfield  {author} {\bibinfo {author} {\bibfnamefont {N.~E.}\ \bibnamefont
  {Hussey}},\ }\href {\doibase 10.1088/0953-8984/20/12/123201} {\bibfield
  {journal} {\bibinfo  {journal} {J. Phys.: Condens. Matter}\ }\textbf
  {\bibinfo {volume} {20}},\ \bibinfo {pages} {123201} (\bibinfo {year}
  {2008})}\BibitemShut {NoStop}%
\bibitem [{\citenamefont {Stewart}(2001)}]{Stewart2001}%
  \BibitemOpen
  \bibfield  {author} {\bibinfo {author} {\bibfnamefont {G.~R.}\ \bibnamefont
  {Stewart}},\ }\href {\doibase 10.1103/revmodphys.73.797} {\bibfield
  {journal} {\bibinfo  {journal} {Rev. Mod. Phys.}\ }\textbf {\bibinfo {volume}
  {73}},\ \bibinfo {pages} {797} (\bibinfo {year} {2001})}\BibitemShut
  {NoStop}%
\bibitem [{\citenamefont {Gunnarsson}\ \emph {et~al.}(2003)\citenamefont
  {Gunnarsson}, \citenamefont {Calandra},\ and\ \citenamefont
  {Han}}]{Gunnarsson2003}%
  \BibitemOpen
  \bibfield  {author} {\bibinfo {author} {\bibfnamefont {O.}~\bibnamefont
  {Gunnarsson}}, \bibinfo {author} {\bibfnamefont {M.}~\bibnamefont
  {Calandra}}, \ and\ \bibinfo {author} {\bibfnamefont {J.~E.}\ \bibnamefont
  {Han}},\ }\href {\doibase 10.1103/revmodphys.75.1085} {\bibfield  {journal}
  {\bibinfo  {journal} {Rev. Mod. Phys.}\ }\textbf {\bibinfo {volume} {75}},\
  \bibinfo {pages} {1085} (\bibinfo {year} {2003})}\BibitemShut {NoStop}%
\bibitem [{\citenamefont {Hartnoll}(2014)}]{Hartnoll2014}%
  \BibitemOpen
  \bibfield  {author} {\bibinfo {author} {\bibfnamefont {S.~A.}\ \bibnamefont
  {Hartnoll}},\ }\href {\doibase 10.1038/nphys3174} {\bibfield  {journal}
  {\bibinfo  {journal} {Nat. Phys.}\ }\textbf {\bibinfo {volume} {11}},\
  \bibinfo {pages} {54} (\bibinfo {year} {2014})}\BibitemShut {NoStop}%
\bibitem [{\citenamefont {Anderson}(2008)}]{Anderson2008}%
  \BibitemOpen
  \bibfield  {author} {\bibinfo {author} {\bibfnamefont {P.~W.}\ \bibnamefont
  {Anderson}},\ }\href {\doibase 10.1103/PhysRevB.78.174505} {\bibfield
  {journal} {\bibinfo  {journal} {Phys. Rev. B}\ }\textbf {\bibinfo {volume}
  {78}},\ \bibinfo {pages} {174505} (\bibinfo {year} {2008})}\BibitemShut
  {NoStop}%
\bibitem [{\citenamefont {Varma}\ \emph {et~al.}(1989)\citenamefont {Varma},
  \citenamefont {Littlewood}, \citenamefont {Schmitt-Rink}, \citenamefont
  {Abrahams},\ and\ \citenamefont {Ruckenstein}}]{Varma1989}%
  \BibitemOpen
  \bibfield  {author} {\bibinfo {author} {\bibfnamefont {C.~M.}\ \bibnamefont
  {Varma}}, \bibinfo {author} {\bibfnamefont {P.~B.}\ \bibnamefont
  {Littlewood}}, \bibinfo {author} {\bibfnamefont {S.}~\bibnamefont
  {Schmitt-Rink}}, \bibinfo {author} {\bibfnamefont {E.}~\bibnamefont
  {Abrahams}}, \ and\ \bibinfo {author} {\bibfnamefont {A.~E.}\ \bibnamefont
  {Ruckenstein}},\ }\href {\doibase 10.1103/physrevlett.63.1996} {\bibfield
  {journal} {\bibinfo  {journal} {Phys. Rev. Lett.}\ }\textbf {\bibinfo
  {volume} {63}},\ \bibinfo {pages} {1996} (\bibinfo {year}
  {1989})}\BibitemShut {NoStop}%
\bibitem [{\citenamefont {Vu\ifmmode \check{c}\else \v{c}\fi{}i\ifmmode
  \check{c}\else \v{c}\fi{}evi\ifmmode~\acute{c}\else \'{c}\fi{}}\ \emph
  {et~al.}(2015)\citenamefont {Vu\ifmmode \check{c}\else \v{c}\fi{}i\ifmmode
  \check{c}\else \v{c}\fi{}evi\ifmmode~\acute{c}\else \'{c}\fi{}},
  \citenamefont {Tanaskovi\ifmmode~\acute{c}\else \'{c}\fi{}}, \citenamefont
  {Rozenberg},\ and\ \citenamefont {Dobrosavljevi\ifmmode~\acute{c}\else
  \'{c}\fi{}}}]{Vucicevic2015}%
  \BibitemOpen
  \bibfield  {author} {\bibinfo {author} {\bibfnamefont {J.}~\bibnamefont
  {Vu\ifmmode \check{c}\else \v{c}\fi{}i\ifmmode \check{c}\else
  \v{c}\fi{}evi\ifmmode~\acute{c}\else \'{c}\fi{}}}, \bibinfo {author}
  {\bibfnamefont {D.}~\bibnamefont {Tanaskovi\ifmmode~\acute{c}\else
  \'{c}\fi{}}}, \bibinfo {author} {\bibfnamefont {M.~J.}\ \bibnamefont
  {Rozenberg}}, \ and\ \bibinfo {author} {\bibfnamefont {V.}~\bibnamefont
  {Dobrosavljevi\ifmmode~\acute{c}\else \'{c}\fi{}}},\ }\href {\doibase
  10.1103/PhysRevLett.114.246402} {\bibfield  {journal} {\bibinfo  {journal}
  {Phys. Rev. Lett.}\ }\textbf {\bibinfo {volume} {114}},\ \bibinfo {pages}
  {246402} (\bibinfo {year} {2015})}\BibitemShut {NoStop}%
\bibitem [{\citenamefont {Hartnoll}\ \emph {et~al.}(2018)\citenamefont
  {Hartnoll}, \citenamefont {Lucas},\ and\ \citenamefont
  {Sachdev}}]{Hartnoll2018}%
  \BibitemOpen
  \bibfield  {author} {\bibinfo {author} {\bibfnamefont {S.~A.}\ \bibnamefont
  {Hartnoll}}, \bibinfo {author} {\bibfnamefont {A.}~\bibnamefont {Lucas}}, \
  and\ \bibinfo {author} {\bibfnamefont {S.}~\bibnamefont {Sachdev}},\ }\href
  {https://www.ebook.de/de/product/31890881/sean_a_hartnoll_andrew_lucas_subir_sachdev_holographic_quantum_matter.html}
  {\emph {\bibinfo {title} {Holographic Quantum Matter}}}\ (\bibinfo
  {publisher} {MIT Press},\ \bibinfo {year} {2018})\BibitemShut {NoStop}%
\bibitem [{\citenamefont {Scalapino}(2007)}]{Scalapino2007}%
  \BibitemOpen
  \bibfield  {author} {\bibinfo {author} {\bibfnamefont {D.~J.}\ \bibnamefont
  {Scalapino}},\ }\enquote {\bibinfo {title} {Handbook of high-temperature
  superconductivity},}\ \ (\bibinfo  {publisher} {Springer New York},\ \bibinfo
  {year} {2007})\ Chap.\ \bibinfo {chapter} {Numerical Studies of the 2D
  Hubbard Model}, pp.\ \bibinfo {pages} {463--493}\BibitemShut {NoStop}%
\bibitem [{\citenamefont {Jakli\ifmmode~\check{c}\else \v{c}\fi{}}\ and\
  \citenamefont {Prelov\ifmmode~\check{s}\else
  \v{s}\fi{}ek}(2000)}]{Jaklic2000}%
  \BibitemOpen
  \bibfield  {author} {\bibinfo {author} {\bibfnamefont {J.}~\bibnamefont
  {Jakli\ifmmode~\check{c}\else \v{c}\fi{}}}\ and\ \bibinfo {author}
  {\bibfnamefont {P.}~\bibnamefont {Prelov\ifmmode~\check{s}\else
  \v{s}\fi{}ek}},\ }\href {\doibase 10.1080/000187300243381} {\bibfield
  {journal} {\bibinfo  {journal} {Adv. Phys.}\ }\textbf {\bibinfo {volume}
  {49}},\ \bibinfo {pages} {1} (\bibinfo {year} {2000})}\BibitemShut {NoStop}%
\bibitem [{\citenamefont {Brantut}\ \emph {et~al.}(2012)\citenamefont
  {Brantut}, \citenamefont {Meineke}, \citenamefont {Stadler}, \citenamefont
  {Krinner},\ and\ \citenamefont {Esslinger}}]{Brantut2012}%
  \BibitemOpen
  \bibfield  {author} {\bibinfo {author} {\bibfnamefont {J.-P.}\ \bibnamefont
  {Brantut}}, \bibinfo {author} {\bibfnamefont {J.}~\bibnamefont {Meineke}},
  \bibinfo {author} {\bibfnamefont {D.}~\bibnamefont {Stadler}}, \bibinfo
  {author} {\bibfnamefont {S.}~\bibnamefont {Krinner}}, \ and\ \bibinfo
  {author} {\bibfnamefont {T.}~\bibnamefont {Esslinger}},\ }\href {\doibase
  10.1126/science.1223175} {\bibfield  {journal} {\bibinfo  {journal}
  {Science}\ }\textbf {\bibinfo {volume} {337}},\ \bibinfo {pages} {1069}
  (\bibinfo {year} {2012})}\BibitemShut {NoStop}%
\bibitem [{\citenamefont {Valtolina}\ \emph {et~al.}(2015)\citenamefont
  {Valtolina}, \citenamefont {Burchianti}, \citenamefont {Amico}, \citenamefont
  {Neri}, \citenamefont {Xhani}, \citenamefont {Seman}, \citenamefont
  {Trombettoni}, \citenamefont {Smerzi}, \citenamefont {Zaccanti},
  \citenamefont {Inguscio},\ and\ \citenamefont {Roati}}]{Valtolina2015}%
  \BibitemOpen
  \bibfield  {author} {\bibinfo {author} {\bibfnamefont {G.}~\bibnamefont
  {Valtolina}}, \bibinfo {author} {\bibfnamefont {A.}~\bibnamefont
  {Burchianti}}, \bibinfo {author} {\bibfnamefont {A.}~\bibnamefont {Amico}},
  \bibinfo {author} {\bibfnamefont {E.}~\bibnamefont {Neri}}, \bibinfo {author}
  {\bibfnamefont {K.}~\bibnamefont {Xhani}}, \bibinfo {author} {\bibfnamefont
  {J.~A.}\ \bibnamefont {Seman}}, \bibinfo {author} {\bibfnamefont
  {A.}~\bibnamefont {Trombettoni}}, \bibinfo {author} {\bibfnamefont
  {A.}~\bibnamefont {Smerzi}}, \bibinfo {author} {\bibfnamefont
  {M.}~\bibnamefont {Zaccanti}}, \bibinfo {author} {\bibfnamefont
  {M.}~\bibnamefont {Inguscio}}, \ and\ \bibinfo {author} {\bibfnamefont
  {G.}~\bibnamefont {Roati}},\ }\href {\doibase 10.1126/science.aac9725}
  {\bibfield  {journal} {\bibinfo  {journal} {Science}\ }\textbf {\bibinfo
  {volume} {350}},\ \bibinfo {pages} {1505} (\bibinfo {year}
  {2015})}\BibitemShut {NoStop}%
\bibitem [{\citenamefont {Krinner}\ \emph {et~al.}(2017)\citenamefont
  {Krinner}, \citenamefont {Esslinger},\ and\ \citenamefont
  {Brantut}}]{Krinner2017}%
  \BibitemOpen
  \bibfield  {author} {\bibinfo {author} {\bibfnamefont {S.}~\bibnamefont
  {Krinner}}, \bibinfo {author} {\bibfnamefont {T.}~\bibnamefont {Esslinger}},
  \ and\ \bibinfo {author} {\bibfnamefont {J.-P.}\ \bibnamefont {Brantut}},\
  }\href {\doibase 10.1088/1361-648x/aa74a1} {\bibfield  {journal} {\bibinfo
  {journal} {J. Phys.: Condens. Matter}\ }\textbf {\bibinfo {volume} {29}},\
  \bibinfo {pages} {343003} (\bibinfo {year} {2017})}\BibitemShut {NoStop}%
\bibitem [{\citenamefont {Lebrat}\ \emph {et~al.}(2018)\citenamefont {Lebrat},
  \citenamefont {Gri\ifmmode~\check{s}\else \v{s}\fi{}ins}, \citenamefont
  {Husmann}, \citenamefont {H\"ausler}, \citenamefont {Corman}, \citenamefont
  {Giamarchi}, \citenamefont {Brantut},\ and\ \citenamefont
  {Esslinger}}]{Lebrat2017}%
  \BibitemOpen
  \bibfield  {author} {\bibinfo {author} {\bibfnamefont {M.}~\bibnamefont
  {Lebrat}}, \bibinfo {author} {\bibfnamefont {P.}~\bibnamefont
  {Gri\ifmmode~\check{s}\else \v{s}\fi{}ins}}, \bibinfo {author} {\bibfnamefont
  {D.}~\bibnamefont {Husmann}}, \bibinfo {author} {\bibfnamefont
  {S.}~\bibnamefont {H\"ausler}}, \bibinfo {author} {\bibfnamefont
  {L.}~\bibnamefont {Corman}}, \bibinfo {author} {\bibfnamefont
  {T.}~\bibnamefont {Giamarchi}}, \bibinfo {author} {\bibfnamefont {J.-P.}\
  \bibnamefont {Brantut}}, \ and\ \bibinfo {author} {\bibfnamefont
  {T.}~\bibnamefont {Esslinger}},\ }\href {\doibase 10.1103/PhysRevX.8.011053}
  {\bibfield  {journal} {\bibinfo  {journal} {Phys. Rev. X}\ }\textbf {\bibinfo
  {volume} {8}},\ \bibinfo {pages} {011053} (\bibinfo {year}
  {2018})}\BibitemShut {NoStop}%
\bibitem [{\citenamefont {Ott}\ \emph {et~al.}(2004)\citenamefont {Ott},
  \citenamefont {de~Mirandes}, \citenamefont {Ferlaino}, \citenamefont {Roati},
  \citenamefont {Modugno},\ and\ \citenamefont {Inguscio}}]{Ott2004}%
  \BibitemOpen
  \bibfield  {author} {\bibinfo {author} {\bibfnamefont {H.}~\bibnamefont
  {Ott}}, \bibinfo {author} {\bibfnamefont {E.}~\bibnamefont {de~Mirandes}},
  \bibinfo {author} {\bibfnamefont {F.}~\bibnamefont {Ferlaino}}, \bibinfo
  {author} {\bibfnamefont {G.}~\bibnamefont {Roati}}, \bibinfo {author}
  {\bibfnamefont {G.}~\bibnamefont {Modugno}}, \ and\ \bibinfo {author}
  {\bibfnamefont {M.}~\bibnamefont {Inguscio}},\ }\href {\doibase
  10.1103/PhysRevLett.92.160601} {\bibfield  {journal} {\bibinfo  {journal}
  {Phys. Rev. Lett.}\ }\textbf {\bibinfo {volume} {92}},\ \bibinfo {pages}
  {160601} (\bibinfo {year} {2004})}\BibitemShut {NoStop}%
\bibitem [{\citenamefont {Strohmaier}\ \emph {et~al.}(2007)\citenamefont
  {Strohmaier}, \citenamefont {Takasu}, \citenamefont {G\"unter}, \citenamefont
  {J\"ordens}, \citenamefont {K\"ohl}, \citenamefont {Moritz},\ and\
  \citenamefont {Esslinger}}]{Strohmaier2007}%
  \BibitemOpen
  \bibfield  {author} {\bibinfo {author} {\bibfnamefont {N.}~\bibnamefont
  {Strohmaier}}, \bibinfo {author} {\bibfnamefont {Y.}~\bibnamefont {Takasu}},
  \bibinfo {author} {\bibfnamefont {K.}~\bibnamefont {G\"unter}}, \bibinfo
  {author} {\bibfnamefont {R.}~\bibnamefont {J\"ordens}}, \bibinfo {author}
  {\bibfnamefont {M.}~\bibnamefont {K\"ohl}}, \bibinfo {author} {\bibfnamefont
  {H.}~\bibnamefont {Moritz}}, \ and\ \bibinfo {author} {\bibfnamefont
  {T.}~\bibnamefont {Esslinger}},\ }\href {\doibase
  10.1103/PhysRevLett.99.220601} {\bibfield  {journal} {\bibinfo  {journal}
  {Phys. Rev. Lett.}\ }\textbf {\bibinfo {volume} {99}},\ \bibinfo {pages}
  {220601} (\bibinfo {year} {2007})}\BibitemShut {NoStop}%
\bibitem [{\citenamefont {Schneider}\ \emph {et~al.}(2012)\citenamefont
  {Schneider}, \citenamefont {Hackerm\"uller}, \citenamefont {Ronzheimer},
  \citenamefont {Will}, \citenamefont {Braun}, \citenamefont {Best},
  \citenamefont {Bloch}, \citenamefont {Demler}, \citenamefont {Mandt},
  \citenamefont {Rasch},\ and\ \citenamefont {Rosch}}]{Schneider2012}%
  \BibitemOpen
  \bibfield  {author} {\bibinfo {author} {\bibfnamefont {U.}~\bibnamefont
  {Schneider}}, \bibinfo {author} {\bibfnamefont {L.}~\bibnamefont
  {Hackerm\"uller}}, \bibinfo {author} {\bibfnamefont {J.~P.}\ \bibnamefont
  {Ronzheimer}}, \bibinfo {author} {\bibfnamefont {S.}~\bibnamefont {Will}},
  \bibinfo {author} {\bibfnamefont {S.}~\bibnamefont {Braun}}, \bibinfo
  {author} {\bibfnamefont {T.}~\bibnamefont {Best}}, \bibinfo {author}
  {\bibfnamefont {I.}~\bibnamefont {Bloch}}, \bibinfo {author} {\bibfnamefont
  {E.}~\bibnamefont {Demler}}, \bibinfo {author} {\bibfnamefont
  {S.}~\bibnamefont {Mandt}}, \bibinfo {author} {\bibfnamefont
  {D.}~\bibnamefont {Rasch}}, \ and\ \bibinfo {author} {\bibfnamefont
  {A.}~\bibnamefont {Rosch}},\ }\href {\doibase 10.1038/nphys2205} {\bibfield
  {journal} {\bibinfo  {journal} {Nat. Phys.}\ }\textbf {\bibinfo {volume}
  {8}},\ \bibinfo {pages} {213} (\bibinfo {year} {2012})}\BibitemShut {NoStop}%
\bibitem [{\citenamefont {Xu}\ \emph {et~al.}()\citenamefont {Xu},
  \citenamefont {McGehee}, \citenamefont {Morong},\ and\ \citenamefont
  {DeMarco}}]{Xu2016}%
  \BibitemOpen
  \bibfield  {author} {\bibinfo {author} {\bibfnamefont {W.}~\bibnamefont
  {Xu}}, \bibinfo {author} {\bibfnamefont {W.}~\bibnamefont {McGehee}},
  \bibinfo {author} {\bibfnamefont {W.}~\bibnamefont {Morong}}, \ and\ \bibinfo
  {author} {\bibfnamefont {B.}~\bibnamefont {DeMarco}},\ }\href@noop {} {\
  }\Eprint {http://arxiv.org/abs/arXiv:1606.06669v4} {arXiv:1606.06669v4}
  \BibitemShut {NoStop}%
\bibitem [{\citenamefont {Anderson}\ \emph {et~al.}()\citenamefont {Anderson},
  \citenamefont {Wang}, \citenamefont {Xu}, \citenamefont {Venu}, \citenamefont
  {Trotzky}, \citenamefont {Chevy},\ and\ \citenamefont
  {Thywissen}}]{Anderson2017}%
  \BibitemOpen
  \bibfield  {author} {\bibinfo {author} {\bibfnamefont {R.}~\bibnamefont
  {Anderson}}, \bibinfo {author} {\bibfnamefont {F.}~\bibnamefont {Wang}},
  \bibinfo {author} {\bibfnamefont {P.}~\bibnamefont {Xu}}, \bibinfo {author}
  {\bibfnamefont {V.}~\bibnamefont {Venu}}, \bibinfo {author} {\bibfnamefont
  {S.}~\bibnamefont {Trotzky}}, \bibinfo {author} {\bibfnamefont
  {F.}~\bibnamefont {Chevy}}, \ and\ \bibinfo {author} {\bibfnamefont {J.~H.}\
  \bibnamefont {Thywissen}},\ }\href@noop {} {\ }\Eprint
  {http://arxiv.org/abs/arXiv:1712.09965v1} {arXiv:1712.09965v1} \BibitemShut
  {NoStop}%
\bibitem [{sup()}]{supplement}%
  \BibitemOpen
  \href@noop {} {}\bibinfo {note} {Materials and methods are available as
  supplementary materials}\BibitemShut {NoStop}%
\bibitem [{\citenamefont {Brown}\ \emph {et~al.}(2017)\citenamefont {Brown},
  \citenamefont {Mitra}, \citenamefont {Guardado-Sanchez}, \citenamefont
  {Schau{\ss}}, \citenamefont {Kondov}, \citenamefont {Khatami}, \citenamefont
  {Paiva}, \citenamefont {Trivedi}, \citenamefont {Huse},\ and\ \citenamefont
  {Bakr}}]{Brown2017}%
  \BibitemOpen
  \bibfield  {author} {\bibinfo {author} {\bibfnamefont {P.~T.}\ \bibnamefont
  {Brown}}, \bibinfo {author} {\bibfnamefont {D.}~\bibnamefont {Mitra}},
  \bibinfo {author} {\bibfnamefont {E.}~\bibnamefont {Guardado-Sanchez}},
  \bibinfo {author} {\bibfnamefont {P.}~\bibnamefont {Schau{\ss}}}, \bibinfo
  {author} {\bibfnamefont {S.~S.}\ \bibnamefont {Kondov}}, \bibinfo {author}
  {\bibfnamefont {E.}~\bibnamefont {Khatami}}, \bibinfo {author} {\bibfnamefont
  {T.}~\bibnamefont {Paiva}}, \bibinfo {author} {\bibfnamefont
  {N.}~\bibnamefont {Trivedi}}, \bibinfo {author} {\bibfnamefont {D.~A.}\
  \bibnamefont {Huse}}, \ and\ \bibinfo {author} {\bibfnamefont {W.~S.}\
  \bibnamefont {Bakr}},\ }\href {\doibase 10.1126/science.aam7838} {\bibfield
  {journal} {\bibinfo  {journal} {Science}\ }\textbf {\bibinfo {volume}
  {357}},\ \bibinfo {pages} {1385} (\bibinfo {year} {2017})}\BibitemShut
  {NoStop}%
\bibitem [{\citenamefont {Vidhyadhiraja}\ \emph {et~al.}(2009)\citenamefont
  {Vidhyadhiraja}, \citenamefont {Macridin}, \citenamefont
  {\ifmmode~\mbox{\c{S}}\else \c{S}\fi{}en}, \citenamefont {Jarrell},\ and\
  \citenamefont {Ma}}]{Vidhyadhiraja2009}%
  \BibitemOpen
  \bibfield  {author} {\bibinfo {author} {\bibfnamefont {N.~S.}\ \bibnamefont
  {Vidhyadhiraja}}, \bibinfo {author} {\bibfnamefont {A.}~\bibnamefont
  {Macridin}}, \bibinfo {author} {\bibfnamefont {C.}~\bibnamefont
  {\ifmmode~\mbox{\c{S}}\else \c{S}\fi{}en}}, \bibinfo {author} {\bibfnamefont
  {M.}~\bibnamefont {Jarrell}}, \ and\ \bibinfo {author} {\bibfnamefont
  {M.}~\bibnamefont {Ma}},\ }\href {\doibase 10.1103/PhysRevLett.102.206407}
  {\bibfield  {journal} {\bibinfo  {journal} {Phys. Rev. Lett.}\ }\textbf
  {\bibinfo {volume} {102}},\ \bibinfo {pages} {206407} (\bibinfo {year}
  {2009})}\BibitemShut {NoStop}%
\bibitem [{\citenamefont {Khatami}\ and\ \citenamefont
  {Rigol}(2011)}]{Khatami2011}%
  \BibitemOpen
  \bibfield  {author} {\bibinfo {author} {\bibfnamefont {E.}~\bibnamefont
  {Khatami}}\ and\ \bibinfo {author} {\bibfnamefont {M.}~\bibnamefont
  {Rigol}},\ }\href {\doibase 10.1103/PhysRevA.84.053611} {\bibfield  {journal}
  {\bibinfo  {journal} {Phys. Rev. A}\ }\textbf {\bibinfo {volume} {84}},\
  \bibinfo {pages} {053611} (\bibinfo {year} {2011})}\BibitemShut {NoStop}%
\bibitem [{\citenamefont {Hild}\ \emph {et~al.}(2014)\citenamefont {Hild},
  \citenamefont {Fukuhara}, \citenamefont {Schau\ss{}}, \citenamefont {Zeiher},
  \citenamefont {Knap}, \citenamefont {Demler}, \citenamefont {Bloch},\ and\
  \citenamefont {Gross}}]{Hild2014}%
  \BibitemOpen
  \bibfield  {author} {\bibinfo {author} {\bibfnamefont {S.}~\bibnamefont
  {Hild}}, \bibinfo {author} {\bibfnamefont {T.}~\bibnamefont {Fukuhara}},
  \bibinfo {author} {\bibfnamefont {P.}~\bibnamefont {Schau\ss{}}}, \bibinfo
  {author} {\bibfnamefont {J.}~\bibnamefont {Zeiher}}, \bibinfo {author}
  {\bibfnamefont {M.}~\bibnamefont {Knap}}, \bibinfo {author} {\bibfnamefont
  {E.}~\bibnamefont {Demler}}, \bibinfo {author} {\bibfnamefont
  {I.}~\bibnamefont {Bloch}}, \ and\ \bibinfo {author} {\bibfnamefont
  {C.}~\bibnamefont {Gross}},\ }\href {\doibase 10.1103/PhysRevLett.113.147205}
  {\bibfield  {journal} {\bibinfo  {journal} {Phys. Rev. Lett.}\ }\textbf
  {\bibinfo {volume} {113}},\ \bibinfo {pages} {147205} (\bibinfo {year}
  {2014})}\BibitemShut {NoStop}%
\bibitem [{\citenamefont {P{\'{a}}lsson}\ and\ \citenamefont
  {Kotliar}(1998)}]{Palsson1998}%
  \BibitemOpen
  \bibfield  {author} {\bibinfo {author} {\bibfnamefont {G.}~\bibnamefont
  {P{\'{a}}lsson}}\ and\ \bibinfo {author} {\bibfnamefont {G.}~\bibnamefont
  {Kotliar}},\ }\href {\doibase 10.1103/physrevlett.80.4775} {\bibfield
  {journal} {\bibinfo  {journal} {Phys. Rev. Lett.}\ }\textbf {\bibinfo
  {volume} {80}},\ \bibinfo {pages} {4775} (\bibinfo {year}
  {1998})}\BibitemShut {NoStop}%
\bibitem [{\citenamefont {Perepelitsky}\ \emph {et~al.}(2016)\citenamefont
  {Perepelitsky}, \citenamefont {Galatas}, \citenamefont {Mravlje},
  \citenamefont {\ifmmode~\check{Z}\else \v{Z}\fi{}itko}, \citenamefont
  {Khatami}, \citenamefont {Shastry},\ and\ \citenamefont
  {Georges}}]{Perepelitsky2016}%
  \BibitemOpen
  \bibfield  {author} {\bibinfo {author} {\bibfnamefont {E.}~\bibnamefont
  {Perepelitsky}}, \bibinfo {author} {\bibfnamefont {A.}~\bibnamefont
  {Galatas}}, \bibinfo {author} {\bibfnamefont {J.}~\bibnamefont {Mravlje}},
  \bibinfo {author} {\bibfnamefont {R.}~\bibnamefont {\ifmmode~\check{Z}\else
  \v{Z}\fi{}itko}}, \bibinfo {author} {\bibfnamefont {E.}~\bibnamefont
  {Khatami}}, \bibinfo {author} {\bibfnamefont {B.~S.}\ \bibnamefont
  {Shastry}}, \ and\ \bibinfo {author} {\bibfnamefont {A.}~\bibnamefont
  {Georges}},\ }\href {\doibase 10.1103/PhysRevB.94.235115} {\bibfield
  {journal} {\bibinfo  {journal} {Phys. Rev. B}\ }\textbf {\bibinfo {volume}
  {94}},\ \bibinfo {pages} {235115} (\bibinfo {year} {2016})}\BibitemShut
  {NoStop}%
\bibitem [{\citenamefont {Kokalj}(2017)}]{Kokalj2017}%
  \BibitemOpen
  \bibfield  {author} {\bibinfo {author} {\bibfnamefont {J.}~\bibnamefont
  {Kokalj}},\ }\href {\doibase 10.1103/PhysRevB.95.041110} {\bibfield
  {journal} {\bibinfo  {journal} {Phys. Rev. B}\ }\textbf {\bibinfo {volume}
  {95}},\ \bibinfo {pages} {041110} (\bibinfo {year} {2017})}\BibitemShut
  {NoStop}%
\bibitem [{\citenamefont {Mukerjee}\ \emph {et~al.}(2006)\citenamefont
  {Mukerjee}, \citenamefont {Oganesyan},\ and\ \citenamefont
  {Huse}}]{Mukerjee2006}%
  \BibitemOpen
  \bibfield  {author} {\bibinfo {author} {\bibfnamefont {S.}~\bibnamefont
  {Mukerjee}}, \bibinfo {author} {\bibfnamefont {V.}~\bibnamefont {Oganesyan}},
  \ and\ \bibinfo {author} {\bibfnamefont {D.}~\bibnamefont {Huse}},\ }\href
  {\doibase 10.1103/PhysRevB.73.035113} {\bibfield  {journal} {\bibinfo
  {journal} {Phys. Rev. B}\ }\textbf {\bibinfo {volume} {73}},\ \bibinfo
  {pages} {035113} (\bibinfo {year} {2006})}\BibitemShut {NoStop}%
\bibitem [{\citenamefont {Leviatan}\ \emph {et~al.}()\citenamefont {Leviatan},
  \citenamefont {Pollmann}, \citenamefont {Bardarson}, \citenamefont {Huse},\
  and\ \citenamefont {Altman}}]{Leviatan2017}%
  \BibitemOpen
  \bibfield  {author} {\bibinfo {author} {\bibfnamefont {E.}~\bibnamefont
  {Leviatan}}, \bibinfo {author} {\bibfnamefont {F.}~\bibnamefont {Pollmann}},
  \bibinfo {author} {\bibfnamefont {J.~H.}\ \bibnamefont {Bardarson}}, \bibinfo
  {author} {\bibfnamefont {D.~A.}\ \bibnamefont {Huse}}, \ and\ \bibinfo
  {author} {\bibfnamefont {E.}~\bibnamefont {Altman}},\ }\href@noop {} {\
  }\Eprint {http://arxiv.org/abs/arXiv:1702.08894v2} {arXiv:1702.08894v2}
  \BibitemShut {NoStop}%
\bibitem [{\citenamefont {Ho}\ and\ \citenamefont {Zhou}(2009)}]{Ho2009}%
  \BibitemOpen
  \bibfield  {author} {\bibinfo {author} {\bibfnamefont {T.-L.}\ \bibnamefont
  {Ho}}\ and\ \bibinfo {author} {\bibfnamefont {Q.}~\bibnamefont {Zhou}},\
  }\href {\doibase 10.1038/nphys1477} {\bibfield  {journal} {\bibinfo
  {journal} {Nat. Phys.}\ }\textbf {\bibinfo {volume} {6}},\ \bibinfo {pages}
  {131} (\bibinfo {year} {2009})}\BibitemShut {NoStop}%
\bibitem [{\citenamefont {Cocchi}\ \emph {et~al.}(2016)\citenamefont {Cocchi},
  \citenamefont {Miller}, \citenamefont {Drewes}, \citenamefont {Koschorreck},
  \citenamefont {Pertot}, \citenamefont {Brennecke},\ and\ \citenamefont
  {K\"ohl}}]{Cocchi2016}%
  \BibitemOpen
  \bibfield  {author} {\bibinfo {author} {\bibfnamefont {E.}~\bibnamefont
  {Cocchi}}, \bibinfo {author} {\bibfnamefont {L.~A.}\ \bibnamefont {Miller}},
  \bibinfo {author} {\bibfnamefont {J.~H.}\ \bibnamefont {Drewes}}, \bibinfo
  {author} {\bibfnamefont {M.}~\bibnamefont {Koschorreck}}, \bibinfo {author}
  {\bibfnamefont {D.}~\bibnamefont {Pertot}}, \bibinfo {author} {\bibfnamefont
  {F.}~\bibnamefont {Brennecke}}, \ and\ \bibinfo {author} {\bibfnamefont
  {M.}~\bibnamefont {K\"ohl}},\ }\href {\doibase
  10.1103/PhysRevLett.116.175301} {\bibfield  {journal} {\bibinfo  {journal}
  {Phys. Rev. Lett.}\ }\textbf {\bibinfo {volume} {116}},\ \bibinfo {pages}
  {175301} (\bibinfo {year} {2016})}\BibitemShut {NoStop}%
\bibitem [{\citenamefont {Deng}\ \emph {et~al.}(2013)\citenamefont {Deng},
  \citenamefont {Mravlje}, \citenamefont {\ifmmode~\check{Z}\else
  \v{Z}\fi{}itko}, \citenamefont {Ferrero}, \citenamefont {Kotliar},\ and\
  \citenamefont {Georges}}]{Deng2013}%
  \BibitemOpen
  \bibfield  {author} {\bibinfo {author} {\bibfnamefont {X.}~\bibnamefont
  {Deng}}, \bibinfo {author} {\bibfnamefont {J.}~\bibnamefont {Mravlje}},
  \bibinfo {author} {\bibfnamefont {R.}~\bibnamefont {\ifmmode~\check{Z}\else
  \v{Z}\fi{}itko}}, \bibinfo {author} {\bibfnamefont {M.}~\bibnamefont
  {Ferrero}}, \bibinfo {author} {\bibfnamefont {G.}~\bibnamefont {Kotliar}}, \
  and\ \bibinfo {author} {\bibfnamefont {A.}~\bibnamefont {Georges}},\ }\href
  {\doibase 10.1103/PhysRevLett.110.086401} {\bibfield  {journal} {\bibinfo
  {journal} {Phys. Rev. Lett.}\ }\textbf {\bibinfo {volume} {110}},\ \bibinfo
  {pages} {086401} (\bibinfo {year} {2013})}\BibitemShut {NoStop}%
\bibitem [{\citenamefont {Xu}\ \emph {et~al.}(2013)\citenamefont {Xu},
  \citenamefont {Haule},\ and\ \citenamefont {Kotliar}}]{Xu2013}%
  \BibitemOpen
  \bibfield  {author} {\bibinfo {author} {\bibfnamefont {W.}~\bibnamefont
  {Xu}}, \bibinfo {author} {\bibfnamefont {K.}~\bibnamefont {Haule}}, \ and\
  \bibinfo {author} {\bibfnamefont {G.}~\bibnamefont {Kotliar}},\ }\href
  {\doibase 10.1103/PhysRevLett.111.036401} {\bibfield  {journal} {\bibinfo
  {journal} {Phys. Rev. Lett.}\ }\textbf {\bibinfo {volume} {111}},\ \bibinfo
  {pages} {036401} (\bibinfo {year} {2013})}\BibitemShut {NoStop}%
\bibitem [{\citenamefont {Kotliar}\ \emph {et~al.}(2006)\citenamefont
  {Kotliar}, \citenamefont {Savrasov}, \citenamefont {Haule}, \citenamefont
  {Oudovenko}, \citenamefont {Parcollet},\ and\ \citenamefont
  {Marianetti}}]{Kotliar2006}%
  \BibitemOpen
  \bibfield  {author} {\bibinfo {author} {\bibfnamefont {G.}~\bibnamefont
  {Kotliar}}, \bibinfo {author} {\bibfnamefont {S.~Y.}\ \bibnamefont
  {Savrasov}}, \bibinfo {author} {\bibfnamefont {K.}~\bibnamefont {Haule}},
  \bibinfo {author} {\bibfnamefont {V.~S.}\ \bibnamefont {Oudovenko}}, \bibinfo
  {author} {\bibfnamefont {O.}~\bibnamefont {Parcollet}}, \ and\ \bibinfo
  {author} {\bibfnamefont {C.~A.}\ \bibnamefont {Marianetti}},\ }\href
  {\doibase 10.1103/revmodphys.78.865} {\bibfield  {journal} {\bibinfo
  {journal} {Rev. Mod. Phys.}\ }\textbf {\bibinfo {volume} {78}},\ \bibinfo
  {pages} {865} (\bibinfo {year} {2006})}\BibitemShut {NoStop}%
\bibitem [{\citenamefont {Stewart}\ \emph {et~al.}(2008)\citenamefont
  {Stewart}, \citenamefont {Gaebler},\ and\ \citenamefont {Jin}}]{Stewart2008}%
  \BibitemOpen
  \bibfield  {author} {\bibinfo {author} {\bibfnamefont {J.~T.}\ \bibnamefont
  {Stewart}}, \bibinfo {author} {\bibfnamefont {J.~P.}\ \bibnamefont
  {Gaebler}}, \ and\ \bibinfo {author} {\bibfnamefont {D.~S.}\ \bibnamefont
  {Jin}},\ }\href {\doibase 10.1038/nature07172} {\bibfield  {journal}
  {\bibinfo  {journal} {Nature}\ }\textbf {\bibinfo {volume} {454}},\ \bibinfo
  {pages} {744} (\bibinfo {year} {2008})}\BibitemShut {NoStop}%
\bibitem [{\citenamefont {Brown}\ \emph {et~al.}(2018)\citenamefont {Brown},
  \citenamefont {Mitra}, \citenamefont {Guardado-Sanchez}, \citenamefont
  {Nourafkan}, \citenamefont {Reymbaut}, \citenamefont {H{\'{e}}bert},
  \citenamefont {Bergeron}, \citenamefont {Tremblay}, \citenamefont {Kokalj},
  \citenamefont {Huse}, \citenamefont {Schau{\ss}},\ and\ \citenamefont
  {Bakr}}]{Brown2018a}%
  \BibitemOpen
  \bibfield  {author} {\bibinfo {author} {\bibfnamefont {P.~T.}\ \bibnamefont
  {Brown}}, \bibinfo {author} {\bibfnamefont {D.}~\bibnamefont {Mitra}},
  \bibinfo {author} {\bibfnamefont {E.}~\bibnamefont {Guardado-Sanchez}},
  \bibinfo {author} {\bibfnamefont {R.}~\bibnamefont {Nourafkan}}, \bibinfo
  {author} {\bibfnamefont {A.}~\bibnamefont {Reymbaut}}, \bibinfo {author}
  {\bibfnamefont {C.-D.}\ \bibnamefont {H{\'{e}}bert}}, \bibinfo {author}
  {\bibfnamefont {S.}~\bibnamefont {Bergeron}}, \bibinfo {author}
  {\bibfnamefont {A.-M.~S.}\ \bibnamefont {Tremblay}}, \bibinfo {author}
  {\bibfnamefont {J.}~\bibnamefont {Kokalj}}, \bibinfo {author} {\bibfnamefont
  {D.~A.}\ \bibnamefont {Huse}}, \bibinfo {author} {\bibfnamefont
  {P.}~\bibnamefont {Schau{\ss}}}, \ and\ \bibinfo {author} {\bibfnamefont
  {W.~S.}\ \bibnamefont {Bakr}},\ }\href {\doibase 10.17605/osf.io/3tf8a}
  {\bibfield  {journal} {\bibinfo  {journal} {Open Science Framework}\ }
  (\bibinfo {year} {2018}),\ 10.17605/osf.io/3tf8a}\BibitemShut {NoStop}%
\bibitem [{\citenamefont {Kokalj}(2018)}]{Kokalj2018}%
  \BibitemOpen
  \bibfield  {author} {\bibinfo {author} {\bibfnamefont {J.}~\bibnamefont
  {Kokalj}},\ } \ \url {https://github.com/jurekokalj/ftlm_hub_cond} (\bibinfo  {publisher}
  {GitHub},\ \bibinfo {year} {2018})\BibitemShut {NoStop}%
\bibitem [{\citenamefont {Hebert}(2018)}]{Hebert2018}%
  \BibitemOpen
  \bibfield  {author} {\bibinfo {author} {\bibfnamefont {C.-D.}~\bibnamefont
  {H{\'{e}}bert}},\ }\url {https://github.com/ZGCDDoo/ctint-science} \ (\bibinfo  {publisher}
  {GitHub},\ \bibinfo {year} {2018})\BibitemShut {NoStop}%
\bibitem [{\citenamefont {Nourafkan}(2018)}]{Nourafkan2018}%
  \BibitemOpen
  \bibfield  {author} {\bibinfo {author} {\bibfnamefont {R.}~\bibnamefont
  {Nourafkan}},\ }\url {https://github.com/RezaNourafkan/Transport-}\ (\bibinfo  {publisher}
  {GitHub},\ \bibinfo {year} {2018})\BibitemShut {NoStop}%
\bibitem [{\citenamefont {Floyd}\ and\ \citenamefont
  {Steinberg}(1976)}]{Floyd1976}%
  \BibitemOpen
  \bibfield  {author} {\bibinfo {author} {\bibfnamefont {R.~W.}\ \bibnamefont
  {Floyd}}\ and\ \bibinfo {author} {\bibfnamefont {L.}~\bibnamefont
  {Steinberg}},\ }\href@noop {} {\bibfield  {journal} {\bibinfo  {journal}
  {Proc. Soc. Inf. Displays}\ }\textbf {\bibinfo {volume} {17}},\ \bibinfo
  {pages} {75} (\bibinfo {year} {1976})}\BibitemShut {NoStop}%
\bibitem [{\citenamefont {Kadanoff}\ and\ \citenamefont
  {Martin}(1963)}]{Kadanoff1963}%
  \BibitemOpen
  \bibfield  {author} {\bibinfo {author} {\bibfnamefont {L.~P.}\ \bibnamefont
  {Kadanoff}}\ and\ \bibinfo {author} {\bibfnamefont {P.~C.}\ \bibnamefont
  {Martin}},\ }\href {\doibase 10.1016/0003-4916(63)90078-2} {\bibfield
  {journal} {\bibinfo  {journal} {Ann. Phys.}\ }\textbf {\bibinfo {volume}
  {24}},\ \bibinfo {pages} {419} (\bibinfo {year} {1963})}\BibitemShut
  {NoStop}%
\bibitem [{\citenamefont {Shastry}(2009)}]{Shastry2009}%
  \BibitemOpen
  \bibfield  {author} {\bibinfo {author} {\bibfnamefont {B.~S.}\ \bibnamefont
  {Shastry}},\ }\href {http://stacks.iop.org/0034-4885/72/i=1/a=016501}
  {\bibfield  {journal} {\bibinfo  {journal} {Rep. Prog. Phys.}\ }\textbf
  {\bibinfo {volume} {72}},\ \bibinfo {pages} {016501} (\bibinfo {year}
  {2009})}\BibitemShut {NoStop}%
\bibitem [{\citenamefont {Peterson}\ and\ \citenamefont
  {Shastry}(2010)}]{Peterson2010}%
  \BibitemOpen
  \bibfield  {author} {\bibinfo {author} {\bibfnamefont {M.~R.}\ \bibnamefont
  {Peterson}}\ and\ \bibinfo {author} {\bibfnamefont {B.~S.}\ \bibnamefont
  {Shastry}},\ }\href {\doibase 10.1103/PhysRevB.82.195105} {\bibfield
  {journal} {\bibinfo  {journal} {Phys. Rev. B}\ }\textbf {\bibinfo {volume}
  {82}},\ \bibinfo {pages} {195105} (\bibinfo {year} {2010})}\BibitemShut
  {NoStop}%
\bibitem [{\citenamefont {Arsenault}\ \emph {et~al.}(2013)\citenamefont
  {Arsenault}, \citenamefont {Shastry}, \citenamefont {S\'emon},\ and\
  \citenamefont {Tremblay}}]{Arsenault2013}%
  \BibitemOpen
  \bibfield  {author} {\bibinfo {author} {\bibfnamefont {L.-F.}\ \bibnamefont
  {Arsenault}}, \bibinfo {author} {\bibfnamefont {B.~S.}\ \bibnamefont
  {Shastry}}, \bibinfo {author} {\bibfnamefont {P.}~\bibnamefont {S\'emon}}, \
  and\ \bibinfo {author} {\bibfnamefont {A.-M.~S.}\ \bibnamefont {Tremblay}},\
  }\href {\doibase 10.1103/PhysRevB.87.035126} {\bibfield  {journal} {\bibinfo
  {journal} {Phys. Rev. B}\ }\textbf {\bibinfo {volume} {87}},\ \bibinfo
  {pages} {035126} (\bibinfo {year} {2013})}\BibitemShut {NoStop}%
\bibitem [{\citenamefont {Kokalj}\ and\ \citenamefont
  {McKenzie}(2015)}]{Kokalj2015}%
  \BibitemOpen
  \bibfield  {author} {\bibinfo {author} {\bibfnamefont {J.}~\bibnamefont
  {Kokalj}}\ and\ \bibinfo {author} {\bibfnamefont {R.~H.}\ \bibnamefont
  {McKenzie}},\ }\href {\doibase 10.1103/PhysRevB.91.205121} {\bibfield
  {journal} {\bibinfo  {journal} {Phys. Rev. B}\ }\textbf {\bibinfo {volume}
  {91}},\ \bibinfo {pages} {205121} (\bibinfo {year} {2015})}\BibitemShut
  {NoStop}%
\bibitem [{\citenamefont {Varney}\ \emph {et~al.}(2009)\citenamefont {Varney},
  \citenamefont {Lee}, \citenamefont {Bai}, \citenamefont {Chiesa},
  \citenamefont {Jarrell},\ and\ \citenamefont {Scalettar}}]{Varney2009}%
  \BibitemOpen
  \bibfield  {author} {\bibinfo {author} {\bibfnamefont {C.~N.}\ \bibnamefont
  {Varney}}, \bibinfo {author} {\bibfnamefont {C.-R.}\ \bibnamefont {Lee}},
  \bibinfo {author} {\bibfnamefont {Z.~J.}\ \bibnamefont {Bai}}, \bibinfo
  {author} {\bibfnamefont {S.}~\bibnamefont {Chiesa}}, \bibinfo {author}
  {\bibfnamefont {M.}~\bibnamefont {Jarrell}}, \ and\ \bibinfo {author}
  {\bibfnamefont {R.~T.}\ \bibnamefont {Scalettar}},\ }\href {\doibase
  10.1103/PhysRevB.80.075116} {\bibfield  {journal} {\bibinfo  {journal} {Phys.
  Rev. B}\ }\textbf {\bibinfo {volume} {80}},\ \bibinfo {pages} {075116}
  (\bibinfo {year} {2009})}\BibitemShut {NoStop}%
\bibitem [{\citenamefont {Kokalj}\ and\ \citenamefont
  {McKenzie}(2013)}]{Kokalj2013}%
  \BibitemOpen
  \bibfield  {author} {\bibinfo {author} {\bibfnamefont {J.}~\bibnamefont
  {Kokalj}}\ and\ \bibinfo {author} {\bibfnamefont {R.~H.}\ \bibnamefont
  {McKenzie}},\ }\href {\doibase 10.1103/PhysRevLett.110.206402} {\bibfield
  {journal} {\bibinfo  {journal} {Phys. Rev. Lett.}\ }\textbf {\bibinfo
  {volume} {110}},\ \bibinfo {pages} {206402} (\bibinfo {year}
  {2013})}\BibitemShut {NoStop}%
\bibitem [{\citenamefont {Poilblanc}(1991)}]{Poilblanc1991}%
  \BibitemOpen
  \bibfield  {author} {\bibinfo {author} {\bibfnamefont {D.}~\bibnamefont
  {Poilblanc}},\ }\href {\doibase 10.1103/physrevb.44.9562} {\bibfield
  {journal} {\bibinfo  {journal} {Phys. Rev. B}\ }\textbf {\bibinfo {volume}
  {44}},\ \bibinfo {pages} {9562} (\bibinfo {year} {1991})}\BibitemShut
  {NoStop}%
\bibitem [{\citenamefont {Georges}\ and\ \citenamefont
  {Kotliar}(1992)}]{Georges1992}%
  \BibitemOpen
  \bibfield  {author} {\bibinfo {author} {\bibfnamefont {A.}~\bibnamefont
  {Georges}}\ and\ \bibinfo {author} {\bibfnamefont {G.}~\bibnamefont
  {Kotliar}},\ }\href {\doibase 10.1103/physrevb.45.6479} {\bibfield  {journal}
  {\bibinfo  {journal} {Phys. Rev. B}\ }\textbf {\bibinfo {volume} {45}},\
  \bibinfo {pages} {6479} (\bibinfo {year} {1992})}\BibitemShut {NoStop}%
\bibitem [{\citenamefont {Kotliar}\ \emph {et~al.}(2001)\citenamefont
  {Kotliar}, \citenamefont {Savrasov}, \citenamefont {P\'alsson},\ and\
  \citenamefont {Biroli}}]{Kotliar2001}%
  \BibitemOpen
  \bibfield  {author} {\bibinfo {author} {\bibfnamefont {G.}~\bibnamefont
  {Kotliar}}, \bibinfo {author} {\bibfnamefont {S.~Y.}\ \bibnamefont
  {Savrasov}}, \bibinfo {author} {\bibfnamefont {G.}~\bibnamefont {P\'alsson}},
  \ and\ \bibinfo {author} {\bibfnamefont {G.}~\bibnamefont {Biroli}},\ }\href
  {\doibase 10.1103/PhysRevLett.87.186401} {\bibfield  {journal} {\bibinfo
  {journal} {Phys. Rev. Lett.}\ }\textbf {\bibinfo {volume} {87}},\ \bibinfo
  {pages} {186401} (\bibinfo {year} {2001})}\BibitemShut {NoStop}%
\bibitem [{\citenamefont {Gull}\ \emph {et~al.}(2008)\citenamefont {Gull},
  \citenamefont {Werner}, \citenamefont {Parcollet},\ and\ \citenamefont
  {Troyer}}]{Gull2008}%
  \BibitemOpen
  \bibfield  {author} {\bibinfo {author} {\bibfnamefont {E.}~\bibnamefont
  {Gull}}, \bibinfo {author} {\bibfnamefont {P.}~\bibnamefont {Werner}},
  \bibinfo {author} {\bibfnamefont {O.}~\bibnamefont {Parcollet}}, \ and\
  \bibinfo {author} {\bibfnamefont {M.}~\bibnamefont {Troyer}},\ }\href
  {\doibase 10.1209/0295-5075/82/57003} {\bibfield  {journal} {\bibinfo
  {journal} {EPL}\ }\textbf {\bibinfo {volume} {82}},\ \bibinfo {pages} {57003}
  (\bibinfo {year} {2008})}\BibitemShut {NoStop}%
\bibitem [{\citenamefont {Werner}\ \emph {et~al.}(2006)\citenamefont {Werner},
  \citenamefont {Comanac}, \citenamefont {de' Medici}, \citenamefont {Troyer},\
  and\ \citenamefont {Millis}}]{Werner2006}%
  \BibitemOpen
  \bibfield  {author} {\bibinfo {author} {\bibfnamefont {P.}~\bibnamefont
  {Werner}}, \bibinfo {author} {\bibfnamefont {A.}~\bibnamefont {Comanac}},
  \bibinfo {author} {\bibfnamefont {L.}~\bibnamefont {de' Medici}}, \bibinfo
  {author} {\bibfnamefont {M.}~\bibnamefont {Troyer}}, \ and\ \bibinfo {author}
  {\bibfnamefont {A.~J.}\ \bibnamefont {Millis}},\ }\href {\doibase
  10.1103/PhysRevLett.97.076405} {\bibfield  {journal} {\bibinfo  {journal}
  {Phys. Rev. Lett.}\ }\textbf {\bibinfo {volume} {97}},\ \bibinfo {pages}
  {076405} (\bibinfo {year} {2006})}\BibitemShut {NoStop}%
\bibitem [{\citenamefont {Werner}\ and\ \citenamefont
  {Millis}(2006)}]{Werner2006a}%
  \BibitemOpen
  \bibfield  {author} {\bibinfo {author} {\bibfnamefont {P.}~\bibnamefont
  {Werner}}\ and\ \bibinfo {author} {\bibfnamefont {A.~J.}\ \bibnamefont
  {Millis}},\ }\href {\doibase 10.1103/PhysRevB.74.155107} {\bibfield
  {journal} {\bibinfo  {journal} {Phys. Rev. B}\ }\textbf {\bibinfo {volume}
  {74}},\ \bibinfo {pages} {155107} (\bibinfo {year} {2006})}\BibitemShut
  {NoStop}%
\bibitem [{\citenamefont {Haule}(2007)}]{Haule2007a}%
  \BibitemOpen
  \bibfield  {author} {\bibinfo {author} {\bibfnamefont {K.}~\bibnamefont
  {Haule}},\ }\href {\doibase 10.1103/PhysRevB.75.155113} {\bibfield  {journal}
  {\bibinfo  {journal} {Phys. Rev. B}\ }\textbf {\bibinfo {volume} {75}},\
  \bibinfo {pages} {155113} (\bibinfo {year} {2007})}\BibitemShut {NoStop}%
\bibitem [{\citenamefont {Gull}\ \emph {et~al.}(2007)\citenamefont {Gull},
  \citenamefont {Werner}, \citenamefont {Millis},\ and\ \citenamefont
  {Troyer}}]{Gull2007}%
  \BibitemOpen
  \bibfield  {author} {\bibinfo {author} {\bibfnamefont {E.}~\bibnamefont
  {Gull}}, \bibinfo {author} {\bibfnamefont {P.}~\bibnamefont {Werner}},
  \bibinfo {author} {\bibfnamefont {A.}~\bibnamefont {Millis}}, \ and\ \bibinfo
  {author} {\bibfnamefont {M.}~\bibnamefont {Troyer}},\ }\href {\doibase
  10.1103/PhysRevB.76.235123} {\bibfield  {journal} {\bibinfo  {journal} {Phys.
  Rev. B}\ }\textbf {\bibinfo {volume} {76}},\ \bibinfo {pages} {235123}
  (\bibinfo {year} {2007})}\BibitemShut {NoStop}%
\bibitem [{\citenamefont {S\'emon}\ \emph {et~al.}(2014)\citenamefont
  {S\'emon}, \citenamefont {Yee}, \citenamefont {Haule},\ and\ \citenamefont
  {Tremblay}}]{Semon2014}%
  \BibitemOpen
  \bibfield  {author} {\bibinfo {author} {\bibfnamefont {P.}~\bibnamefont
  {S\'emon}}, \bibinfo {author} {\bibfnamefont {C.-H.}\ \bibnamefont {Yee}},
  \bibinfo {author} {\bibfnamefont {K.}~\bibnamefont {Haule}}, \ and\ \bibinfo
  {author} {\bibfnamefont {A.-M.~S.}\ \bibnamefont {Tremblay}},\ }\href
  {\doibase 10.1103/PhysRevB.90.075149} {\bibfield  {journal} {\bibinfo
  {journal} {Phys. Rev. B}\ }\textbf {\bibinfo {volume} {90}},\ \bibinfo
  {pages} {075149} (\bibinfo {year} {2014})}\BibitemShut {NoStop}%
\bibitem [{\citenamefont {Caffarel}\ and\ \citenamefont
  {Krauth}(1994)}]{Caffarel1994}%
  \BibitemOpen
  \bibfield  {author} {\bibinfo {author} {\bibfnamefont {M.}~\bibnamefont
  {Caffarel}}\ and\ \bibinfo {author} {\bibfnamefont {W.}~\bibnamefont
  {Krauth}},\ }\href {\doibase 10.1103/PhysRevLett.72.1545} {\bibfield
  {journal} {\bibinfo  {journal} {Phys. Rev. Lett.}\ }\textbf {\bibinfo
  {volume} {72}},\ \bibinfo {pages} {1545} (\bibinfo {year}
  {1994})}\BibitemShut {NoStop}%
\bibitem [{\citenamefont {Arsenault}\ and\ \citenamefont
  {Tremblay}(2013)}]{Arsenault2013a}%
  \BibitemOpen
  \bibfield  {author} {\bibinfo {author} {\bibfnamefont {L.-F.}\ \bibnamefont
  {Arsenault}}\ and\ \bibinfo {author} {\bibfnamefont {A.-M.~S.}\ \bibnamefont
  {Tremblay}},\ }\href {\doibase 10.1103/PhysRevB.88.205109} {\bibfield
  {journal} {\bibinfo  {journal} {Phys. Rev. B}\ }\textbf {\bibinfo {volume}
  {88}},\ \bibinfo {pages} {205109} (\bibinfo {year} {2013})}\BibitemShut
  {NoStop}%
\bibitem [{\citenamefont {Maldague}(1977)}]{Maldague1977}%
  \BibitemOpen
  \bibfield  {author} {\bibinfo {author} {\bibfnamefont {P.~F.}\ \bibnamefont
  {Maldague}},\ }\href {\doibase 10.1103/physrevb.16.2437} {\bibfield
  {journal} {\bibinfo  {journal} {Phys. Rev. B}\ }\textbf {\bibinfo {volume}
  {16}},\ \bibinfo {pages} {2437} (\bibinfo {year} {1977})}\BibitemShut
  {NoStop}%
\bibitem [{\citenamefont {Calandra}\ and\ \citenamefont
  {Gunnarsson}(2003)}]{Calandra2003}%
  \BibitemOpen
  \bibfield  {author} {\bibinfo {author} {\bibfnamefont {M.}~\bibnamefont
  {Calandra}}\ and\ \bibinfo {author} {\bibfnamefont {O.}~\bibnamefont
  {Gunnarsson}},\ }\href {http://stacks.iop.org/0295-5075/61/i=1/a=088}
  {\bibfield  {journal} {\bibinfo  {journal} {EPL}\ }\textbf {\bibinfo {volume}
  {61}},\ \bibinfo {pages} {88} (\bibinfo {year} {2003})}\BibitemShut {NoStop}%
\end{thebibliography}
\end{document}